\documentclass[aps,pra,nofootinbib,preprintnumbers]{revtex4-2}
\usepackage{graphicx}
\usepackage{bbold}
\usepackage{slashed}
\usepackage{amsmath}
\usepackage{amssymb}
\usepackage{booktabs}

\newcommand{\bea}{\begin{eqnarray}}
\newcommand{\eea}{\end{eqnarray}}
\newcommand{\be}{\begin{equation}}
\newcommand{\ee}{\end{equation}}

\usepackage[colorinlistoftodos]{todonotes}
\usepackage{appendix}
\usepackage{hyperref}

\begin{document}


\title{On the Potential of Microtubules for Scalable Quantum Computation}

\vspace{0.5cm}

\author{Nick E. Mavromatos$^{1,2}$}

\author{Andreas Mershin$^{3,4,8}$}

\author{Dimitri V. Nanopoulos$^{5,6,7,9}$}

\vspace{1.5cm}

\affiliation{$^1$ Physics Division, School of Applied Mathematical and Physical Sciences, National Technical University of Athens, Zografou Campus, Athens 157 80, Greece}

\affiliation{$^2$Theoretical Particle Physics and Cosmology Group, Department of Physics, King's College London, London, WC2R 2LS, UK}

\affiliation{$^3$MIT Sloan School of Management, 77 Massachusetts Ave. Massachusetts Institute of Technology, Cambridge, MA 02139, USA}

\affiliation{$^4$ www.RealNose.ai, 626 Massachusetts Ave., 2nd Floor, Arlington, MA 02476 , USA}

\affiliation{$^5$Academy of Athens, Division of Natural Sciences, Athens 10679, Greece}

\affiliation{$^6$George P. and Cynthia W. Mitchell Institute for Fundamental Physics and Astronomy, Texas A \& M University, College Station, TX 77843, USA}

\affiliation{$^7$Theoretical Physics Department, CERN, CH-1211 Geneva 23, Switzerland}

\affiliation{$^8$The Osmocosm Public Benefit Foundation, www.OsmoCosm.org Boston, MA, USA}

\affiliation{$^9$The  Digital Health Literacy \& Policy Hub Foundation, \url{www.digitalhealth-hub.com}, NY, USA}

\begin{abstract}

We examine the quantum coherence properties of tubulin heterodimers arranged into the protofilaments of cytoskeletal microtubules. In the physical model proposed by the authors, the microtubule interiors are treated as high-Q quantum electrodynamics (QED) cavities that can support decoherence-resistant entangled states under physiological conditions, with decoherence times of the order of $\mathcal{O}(10^{-6})$~sec. We identify strong electric dipole interactions between tubulin dimers and ordered water dipole quanta within the microtuble interior as the mechanism responsible for the extended coherence times. Classical nonlinear (pseudospin) $\sigma$-models describing solitonic excitations are reinterpreted as emergent quantum-coherent—or possibly pointer—states, arising from incomplete collapse of dipole-aligned quantum states. These solitons mediate dissipation-free energy transfer along microtubule filaments. We discuss logic-gate-like behavior facilitated by microtubule-associated proteins, and outline how such structures may enable scalable, ambient-temperature quantum computation, with the fundamental unit of information storage realized as a quDit encoded in the tubulin dipole state. We further describe a process akin to "decision making" that emerges following an external stimulus, whereby optimal, energy-loss-free signal and information transport pathways are selected across the microtubular network. Finally, we propose experimental approaches—including Rabi-splitting spectroscopy and entangled surface plasmon probes—to validate the use of biomatter as a substrate for scalable quantum computation.

\end{abstract}

\maketitle

\section{Introduction: 
Solitons, dissipationless energy transfer and quantum entanglement in biosystems}\label{sec:intro}

The role of biomatter in quantum computation remains an open question. This issue is explored in~\cite{mmn1} through a model of cytoskeletal microtubules (MTs)~\cite{mt}, treated as high-quality quantum electrodynamics (QED) cavities~\cite{mn1,nem}, with a particular focus on the efficiency of energy—and therefore information—transduction and transport in biological systems. In our previous work, we argued that the QED cavity model of MTs may be instrumental in understanding memory encoding and retrieval functions~\cite{nano}, as it enables the conversion of external stimulus into corresponding electromagnetic signals. QED cavities are well known for facilitating efficient manipulation of quantum entanglement between atoms and photons~\cite{haroche}. In an analogous manner, we proposed in~\cite{mn1} that the electric dipole quanta associated with tubulin dimer walls in the MT cavity interior can become entangled with the dipole quanta of the ordered water in the MT core, giving rise to solitonic dipole structures that mediate dissipation-free energy and signal transport.

Microtubule (MT) dipole systems can be modeled using condensed-matter approaches, such as pseudospin nonlinear 
$\sigma$-models~\cite{mexico}, allowing loss-free signal transduction to be quantified via explicit solitonic solutions to the corresponding equations of motion. As shown in~\cite{mn1,nem,mmn1} and expanded upon here, these classical solitons may be viewed as coherent states of the underlying dipole quanta.\footnote{\label{foot1}  MTs are ubiquitous structures in all eukaryotic cells~\cite{mt}, but there are many other biomolecules involved in memory encoding and transmission for instance: recently it has been proved possible to transplant memories affecting behavior carried by RNA molecules (which themselves have been known to carry genetic information for decades ~\cite{rna}, but not actual cognitive memories). Direct learned memory transfer by RNA injection has been shown between snails ~\cite{rnamemory}, that is, passing down RNA through flesh creates memories in the recipient instead of the conventional path of perceptions translated into neurotransmitters and action potentials through an intact nervous system. Our model for information processing may generally apply to such  biological structures as well but treatment of them is outside the scope of this paper.}

Classical physics allows dissipation-free transport of energy by means of solitons~\cite{solitonsrev}, that are well-established classical field theory configurations. The role of biological solitons in efficient energy and signal transmission, has a long, distinguished history going back to seminal publications by H. Fr\"ohlich~\cite{froehlich} and  A.S. Davydov~\cite{davydov}.  The former showed that observable effects of quantum coherent phenomena in biological systems can occur through coherent excitations in the microwave region due to nonlinear couplings between biomolecular dipoles, leading to solitonic configurations responsible for loss-free energy (i.e.  dissipationless) signaling. The 'pumping' frequency of such coherent modes was established to be of order of the inverse of Fr\"ohlich's coherence time $t_{\rm coherence~Froehlich} \sim 10^{-11}- 10^{-12}~{\rm s}.$ 

Davydov\cite{davydov}, subsequently proposed the existence of \emph{solitonic excitation states} along the $\alpha$-helix self-trapped amide treatment bearing a striking mathematical similarity to superconductivity. The $\alpha$-helix lattice is characterised by two kinds of excitations: deformational oscillations, which result in quantized excitations similar to ``phonons'' in the case of superconductivity, and internal amide excitations. The non-linear coupling between these two types of excitations gives rise to a soliton, which traps the vibrational energy of the $\alpha$-helix, thereby preventing its distortion, and thus resulting in dissipation-free energy transport.

It should be stressed that, although the solitons appear as classical solutions of certain field equations, nonetheless their appearance is the explicit result of quantum coherent states.  
However, due to the extremely complex environment of biological entities, one expects the quantum effects to decohere~\cite{zurek} quite quickly, thus making the conditions for the appearance of coherent states, that could lead to solitons, very delicate, but not impossible, to be realised in nature.\footnote{It should be mentioned, at this point, that the question whether quantum effects play a r\^ole in biophysical systems is much older than the abovementioned works, and dates back to Schr\"odinger~\cite{schr},  who argued that certain aspects of life, such as mutations in living organisms (that is, changes in the DNA sequence of a cell's genom or a virus), might not be explainable by classical physics but require quantum concepts, such as quantum leaps.}  

In the 1990's, a suggestion on the role of quantum effects on brain functioning, and in particular on conscious perception, has been put forward by  S.~Hameroff and R.~Penrose (HP)~\cite{ph}, who concentrated on MTs~\cite{mt} of the brain cells. Specifically, by considering the tubulin heterodimers conformation as quantum states of a two-state system, they assumed 
their coherent superposition, which may result in helicoidal solitonic states propagating along the MT,\footnote{\label{foot3} Note that the helix is a geodesic path (optimal ``minimum distance'' path) in the Euclidean cylindrical geometry $S^1 \otimes \mathbb R$, embeddable in $\mathbb R^3$.} and thus being responsible for conscious perception. HP assumed sufficiently long decoherence time for this purpose of order ${\mathcal O}(1~s)$ in {\it in vivo} situations, so that the \emph{in vivo} system of MT in the brain undergoes, as a result of sufficient growth that allowed it to reach a critical mass/energy, \emph{self-collapse} related to a \emph{quantum gravity} environment (orchestrated reduction method), as opposed to the standard \emph{environmental non-gravitational decoherence} that physical quantum systems are subjected to~\cite{zurek}. 

In our discussion below we shall not adopt the approach of HP~\cite{ph} on understanding consciousness, but we shall be dealing with  more mundane questions, as to whether appropriate solitonic states in MT stem from quantum effects, thus behaving as coherent (or even pointer~\cite{zurek2} states, which had not completely decohered) and whether there are units inside a MT network that 
play a crucial r\^ole in `decision' making regarding the optimal path for information and energy transduction.

\begin{figure}[t]
\begin{center}
\includegraphics[width=9cm]{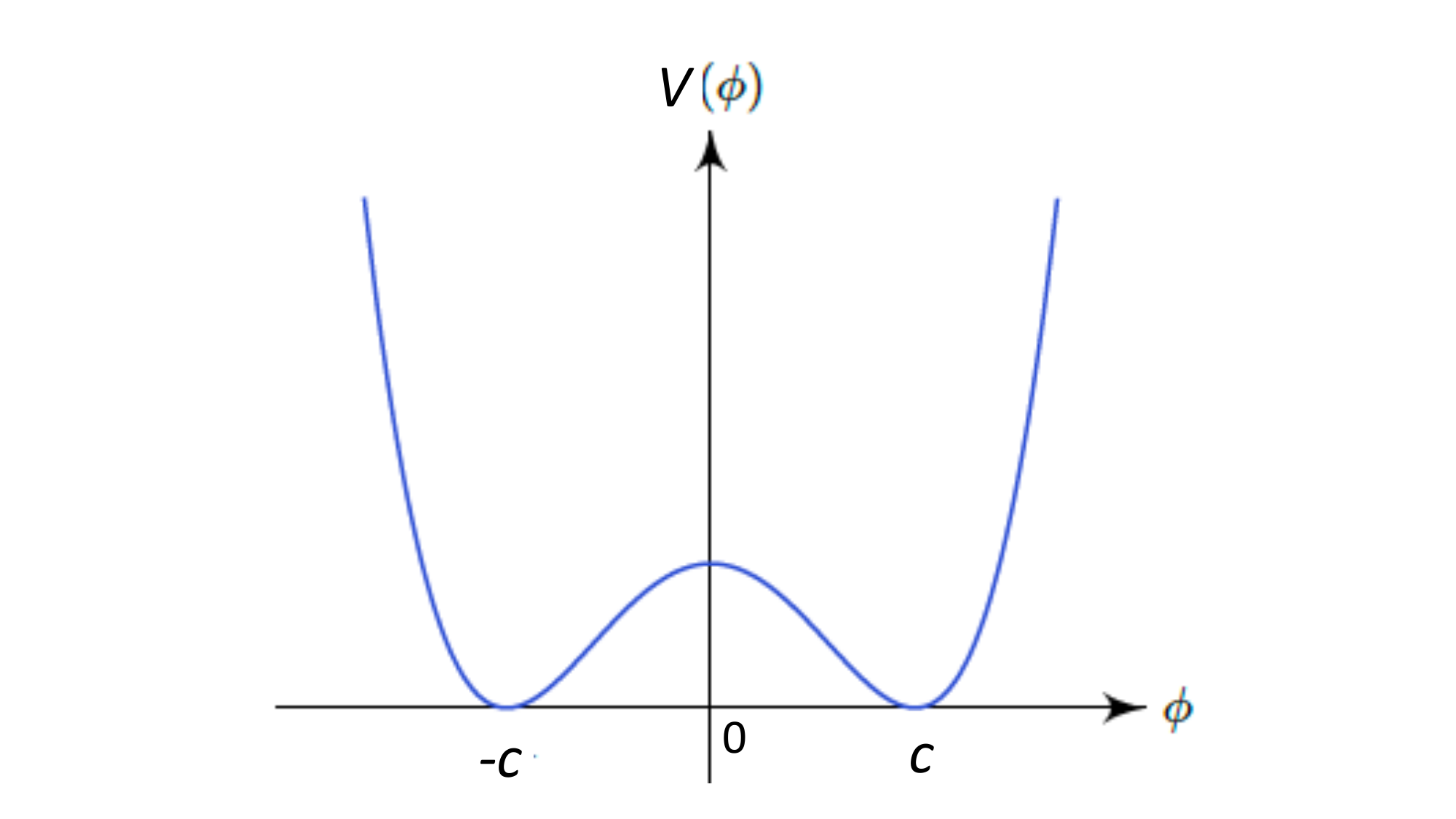}
\end{center}
\caption{A typical potential, with two degenerate non-trivial minima at $\phi=\pm \, C$, 
for a soliton solution in a (1+1)-dimensional real-scalar $\phi$ field theory.}
\label{fig:pot}
\end{figure}

To make the above ideas on the potentially important role of solitons in biological systems clearer to the reader, we  first briefly review the relevant basic properties of  solitons ~\cite{solitonsrev},  in the framework of a toy field-theoretic model. Solitons are {\it finite energy} classical solutions of the Lagrange equations stemming from appropriate field theories. Given the properties of solitons salient to dissipationless energy and signal transduction, it is sufficient (but also directly relevant to the case of biological systems such as microtubular networks), to consider a flat-spacetime (1+1)-dimensional field theory of an interacting scalar field $\phi(t,x)$ with potential $V(\phi)$ and action (for our discussion of soliton solutions below, we use units $\hbar=c=1$, for convenience):
\begin{align}\label{solaction}
 \mathcal S = \int dt\, dx \Big(\frac{1}{2} (\partial_t \phi)^2 - \frac{1}{2} (\partial_x \phi)^2  - V(\phi)\Big)\,.  
\end{align}
We consider potentials with two non-trivial minima, at $\phi = \pm \, C$, as in fig.~\ref{fig:pot}. A typical form is given by:
\begin{align}\label{pot}
   V(\Phi) &= V_0\, (C^2 - \phi^2)^2\,,\nonumber \\
 {\rm where~e.g.} \quad V_0&=\frac{\lambda}{4}\,, \,\, C^2=\frac{m^2}{\lambda} \,, \,\, m>0\,, \lambda > 0\,.   
\end{align}
The total energy functional is given by the spatial integral:
\begin{align}\label{energy}
 E = \int dx \, \Big(\frac{1}{2}(\partial_t \phi)^2  + \frac{1}{2} (\partial_x \phi)^2  + V(\phi)\Big)\,.   
\end{align}
The Euler-Lagrange equations of motion (EoM) stemming from \eqref{solaction} read:
\begin{align}\label{eom}
\partial_t^2\phi - \partial_x^2 \phi = - \frac{\delta V}{\delta \phi}\,.
\end{align}

Solitons are solutions of \eqref{eom}, for which the energy functional \eqref{energy} is {\it finite}. The important feature of solitons is that they are localised solutions, which retain their shape upon propagating freely, but also under collisions among themselves. It is such properties that play a crucial r\^ole in treating solitons as the enablers of loss-free energy transfer, but also as bio- ``logic gates", as we shall discuss below.

To construct a soliton, we first find a static solution to \eqref{eom}, {\it i.e.} $\partial_t \phi =0$, 
and then perform a Lorentz boost, with velocity $v$ along the $x$-direction. The boost is characterized 
by $\partial_t^2 \phi = 0$. Multiplying the static EoM \eqref{eom} by $\partial_x \phi \equiv \phi^\prime$, we obtain the non-linear equation $0=\phi^{\prime\, \prime} \, \phi^\prime  - \frac{\delta V}{\delta \phi} \, \phi^\prime = \frac{d}{dx} \Big( \frac{1}{2}(\phi^\prime)^2 - V(\phi) \Big)$, which can be 
straightforwardly integrated over $x$, 
yielding (upon setting the appropriate constants to zero):
\begin{align}\label{sol1}
  \frac{1}{2} (\phi^\prime)^2 &=  V(\phi) \quad \Rightarrow \quad 
  \phi^\prime = \pm \sqrt{2V(\phi)} \, \nonumber \\
  x-x_0 &= \pm \int_{\phi(x_0)}^{\phi(x)} \, \frac{d\widetilde \phi}{\sqrt{2V(\widetilde \phi)}}\,.
\end{align}
The $\pm$ signature in the right hand side of the middle and final equations is the origin of the {\it chiral} unintuitive nature of the soliton.
 
\begin{figure}[ht]
\begin{center}
\includegraphics[width=10cm]{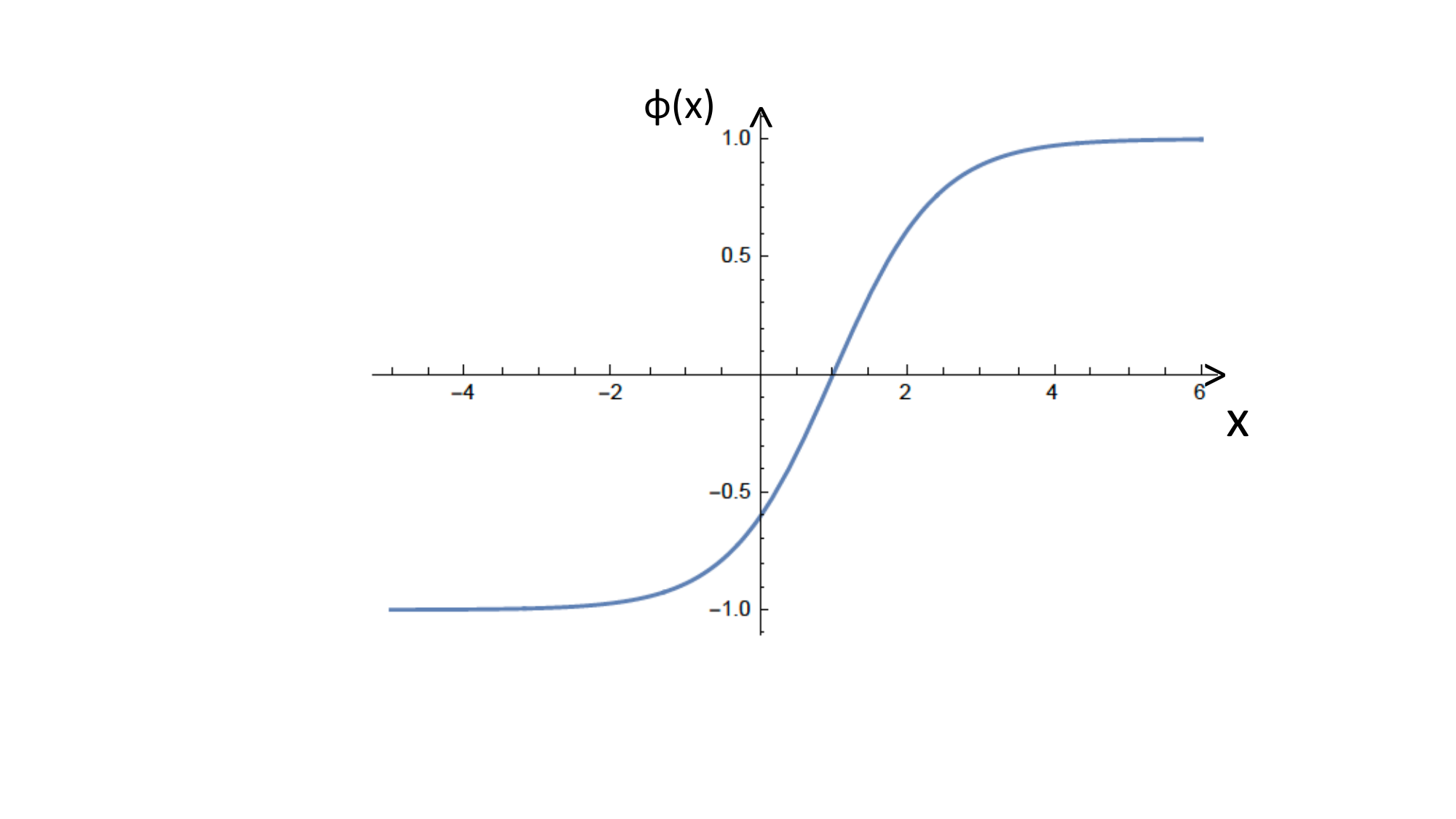}
\end{center}
\vspace{-0.5cm}
\caption{The profile of a typical one-dimensional static kink soliton \eqref{kink} (pictured here for indicative values $m=\sqrt{\lambda}=1$, and $x_0=1$). The same profile characterizes the traveling kink wave \eqref{boostedkink}, obtained upon the replacement of $x-x_0$ in \eqref{kink} by the Lorentz boosted coordinate $\gamma \, (x-x_0 - vt)$, $\gamma = \frac{1}{\sqrt{1-v^2}}$, where $v>0 \, (<0)$ corresponds to left (right) movers.}
\label{fig:kink}
\end{figure}

For specific families of the potential \eqref{pot}, the corresponding soliton solutions are determined by the requirement that the field solutions asymptote (as $x \to \pm \infty$) to the vacuum values:
\begin{align}\label{bcsol}
\lim_{x \to \pm \infty}\phi(x) =  \pm \sqrt{C}= \pm \frac{m}{\sqrt{\lambda}}\,.
\end{align}
 For the $\phi^4$ potential \eqref{pot}, the inversion of the expression \eqref{sol1} yields the celebrated static (anti) kink soliton solution~\cite{solitonsrev}:
 \begin{align}\label{kink}
\phi(x) = \pm \frac{m}{\sqrt{\lambda}}\, {\rm tanh}\Big(\frac{m}{\sqrt{2}}\, (x-x_0)\Big)\,,
\end{align}
which satisfies the boundary conditions \eqref{bcsol} ({\it cf.} fig.~\ref{fig:kink}).  

The energy density $\mathcal E (x)$ of the kink is determined from the integrand of \eqref{energy}, as:
\begin{align}\label{kinkenergy}
    {\mathcal E} (x) = \frac{m^4}{2\,\lambda} \, {\rm sech}^4 \Big(\frac{m}{\sqrt{2}} \, (x -x_0)\Big)\,,
\end{align}
and, therefore, its total mass (total rest energy) is given by:
\begin{align}\label{kinkmass}
  M = \int_{-\infty}^{\infty} \mathcal E (x) \, dx = \frac{2 \sqrt{2}}{3}\, \frac{m^3}{\lambda}  \,,
\end{align}
and it is conserved in time, that is one gets the same value upon using the boosted kink solution (traveling wave), obtained from \eqref{kink} by the replacement $x-x_0 \to \gamma (x-x_0-vt)$, $\gamma = (1-v^2)^{-1/2}$, as already mentioned:
\begin{align}\label{boostedkink}
\phi(t, x) =  \pm \frac{m}{\sqrt{\lambda}}\, {\rm tanh}\Big(\frac{m}{\sqrt{2}}\, \gamma \, (x-x_0 -vt)\Big)\,.
\end{align}

The way the solitons behave at spatial infinity (x = $\pm \infty$ ) imply that it takes an infinite amount of energy to change the (anti)kink configuration. This is an indication of its stability, but the latter is also a guarantee -rigorously for topological reasons. Indeed, for the system \eqref{solaction}, although there are no Noether currents associated with a symmetry of the real scalar field, nevertheless, there is a (trivially) conserved current
\begin{align}\label{conscurr}
\mathcal J^\mu = \frac{1}{2}\,\frac{\sqrt{\lambda}}{m}\, \epsilon^{\mu\nu} \, \partial_\nu \phi\,,
\end{align}
where $\epsilon^{\mu\nu}$ is the Levi-Civita symbol in the (1+1)-dimensional Minkowski spacetime, with the convention $\epsilon^{01}=+1 = - \epsilon^{10}, \, \epsilon^{11}=\epsilon^{00}=0$. The corresponding conserved charge for the soliton (kink) solution 
\begin{align}\label{topcharge}
   Q &= \int_{-\infty}^{\infty} dx \mathcal J^0  
   = \frac{1}{2} 
   \, \frac{\sqrt{\lambda}}{m} \int_{-\infty}^{\infty} dx \, \phi^\prime (x) \nonumber \\
   &= \frac{1}{2} 
   \, \frac{\sqrt{\lambda}}{m} \Big(\phi (+\infty) - \phi(-\infty)\Big)\,.
\end{align}
The above relation \eqref{topcharge} implies that a constant $\phi$ solution is, as expected, characterised by a trivial $Q=0$, but the kink soliton solution is characterised by a non-trivial charge $Q=+1$  ($Q=-1)$ for the anti-kink).  
 
Because $Q$ is a constant, the kinks are stable, and never decay to a solution with $Q=0$, thus $Q$ plays the role of a {\it topological charge}~\cite{solitonsrev}, and the stability of the kink soliton is due to topological reasons. We mention for completion that the topological charge can provide an equivalence-class classification for the various soliton solutions of a field-theory system. Solutions corresponding to the same value of $Q$ belong to the same equivalence class, despite the fact that they may look different in form. The above considerations characterize all types of solitons, even in higher-dimensional field theoretic systems. 

The topological stability of these solitons posits intriguing consequences for quantum computation. While the framework is general, we focus it here on the specifics of microtubular networks made of tubulin, microtubule associated proteins, GTP and GDP all inside the highly dynamic aqueous milieu of the cytosol literally bathed in thermal noise of the order of  $k_B T= 4.3 \times 10^{-21}$\,J or 27\, meV at 37$^{\rm o}$~C (or 310$^{\rm o}$~K). We demonstrate how in our mathematical framework loss-less soliton configurations occur and therefore relevant to the development  quantum computation out of biologically scalable structures. We use  {\it optimal path} selection (on behalf of the biosystem) as the handle. 

While it is true these solitons appear as classical solutions, nonetheless they are the result,of {\it purely quantum} coherent states, formed in biological systems under physiological conditions.  However, due to the complex environment of typically aqueous microenvironment of biological entities, the long-assumed view has been that any quantum effects  decohere~\cite{zurek} too quickly to matter for anything actually observable and of biological relevance. 

It is true that the conditions necessary for the appearance of coherent states leading to the solitons proposed here are delicately fine-tuned, but that appears to be the unspoken rule of biomolecular interactions in biological systems. From the delicately coordinated balance of enzymatic activity to receptors routinely capable of single-molecule capture and even differentiation between isotopes (a literally {\it subatomic} feature!) via systems as grossly integrated as the olfactory sensing and memory apparatus of {\it Drosophila melanogaster} fruitflies ~\cite{MershinFlies2004, MershinFlies2011} what appears to be incredible odds for a laboratory to achieve can sometimes be commonplace occurrences in biological systems.

At this stage we consider it as important to 
point out that~\cite{nmdice10}, even if a MT decoheres in a time scale as short as a few hundreds of femtoseconds, this still allows for quantum effects to play a significant role associated with a `decision' on behalf of realistic biological systems on the most optimal path for efficient energy and signal transfer. Indeed, short decoherence times of that order have been shown in \cite{collini,algae} to be sufficient for information processing via quantum entanglement at ambient temperatures in \emph{cryptophyte marine algae}.\footnote{We note, for completeness, that experimental demonstration on the role of quantum  effects in biological systems has started becoming available already since since 2007, when research work on photosynthesis in plants~\cite{photo} has presented rather convincing experimental evidence that light-absorbing molecules in some photosynthetic proteins capture and transfer energy according to \emph{quantum-mechanical probability laws} instead of classical laws at temperatures up to 180$^{\rm o}$ K.}

In these algae, there are eight chromophore antennae (that change their absorption characteristics and therefore colour upon absorbing certain wavelengths) held in place by the protein scaffold creating the light harvesting complexes. The electronic absorption~\cite{algae}  spectrum of this complex system, was elucidated by applying a laser pulse of  ~25 fs duration. The pulse excites a coherent superposition (in the form of a wave packet) of the antenna's vibrational-electronic eigenstates. The quantum evolution of such as system of coupled bilin molecules under these initial conditions, predicts that the  excitation subsequently oscillates in time between the positions at which the excitation is localized, with distinct correlations and anti-correlations in phase and amplitude. Such coherent oscillations last until the natural eigenstates are restored due to \emph{decoherence}, as a consequence of environmental entanglement~\cite{zurek}.  These experimental results \cite{algae} confirmed such behaviour, showing  a quantum superposition of the electronic structure of the bilin molecule dimer dihydrobiliverdin at room-temperature system lifetimes of order  
\begin{equation}\label{algaedecoh}
t_{\rm decoh} = 400 ~{\rm fs} = 4 \cdot 10^{-13}~{\rm s}~.
\end{equation}

The quantum oscillations of these molecules were transmitted to the other bilin molecules in the complex, at distances 20 Angstr\"oms apart, "as if these molecules were connected by springs". Others have previously~\cite{algae}, concluded that distant molecules within the photosynthetic light harvesting protein complexes are  {\it "long-range, multipartite quantum entangled even at physiological temperatures"} ~\cite{EntangledLHCNature2010}. Long-lived quantum coherence in the photosynthetic Fenna Matthews-Olson Complex~\cite{FMOBioPhysJ2012} and quantum excitation transport \cite{FMOBioPhysJ2012} show quantum entanglement surviving over biologically relevant distances creating observable effects in living matter at ambient temperature. This explicitly assigns a critical role to the exclusively quantum phenomenon of entanglement as necessary for the observed path optimization of energy transmission. This has clear implications for quantum computation which we explore in Sections IV-VI below.

Similar arrangements for extreme-quantum efficiency energy transduction have been studied in the context of photosynthesis ~\cite{EntangledLHCNature2010} , and have guided attempts at creating artificial leaves \cite{mershinNature2012}.

The list of systems found in nature that exploit the toolkit exclusive to quantum physics has grown to include species and settings of such a wide range that it leads us to wonder why we ever failed to expect evolution would be exploiting the availability of quantum trickery -as it can confer significant adaptive advantage such as for instance lossless energy transduction and path optimized transmission finding the most efficient paths over distances of the order of a few nm, spanning several typical protein lengths and highly relevant to receptors and other small-molecule binding biostructures. 

The structure of this article is the following: in the next section \ref{sec:MTqm}, we review the main theoretical pseudospin model of MT~\cite{mexico}, which admits various solitonic solutions, among which helicoidal snoidal waves, which will play an important r\^ole in our analysis, proving a crucial feature for the r\^ole of MT as quantum (bio)computers. Such a classical pseudispin model is viewed here as a result of a (partial) collapse of a the newtwork of (quantum) tubulin dimr states, as we explain in some detail. In section \ref{sec:MTgates} we review the r\^ole of networks of MTs as classical logic gates, giving emphasis on the r\^ole of solitons, as well as the Microtubule associated proteins (MAP), which provide a crucial connection across different MTs in the newtwork, thus enforcing their r\^ole as logic gates. In section \ref{sec:qucom} we describe the basic ingredients for the MTs to operate as scalable biocomputers, namely we discuss the basic information storage unit, the relevant quantum-decoherence mechanism, and the ``decision-making'' process on the most effective path to be followed for a dissipation-free signal and information transduction across the MT. All the above processes take place within the quantum-decoherence time. We provide a microscopic mechanism, within the QED Cavity model of MT~\cite{mn1,nem,mmn1}, which ensures a relatively long decoherence-time interval, allowing for crucial biocomputing processes to take place. The important r\^ole of the ordered-water interia of the MT in this respect is highlighted.
In section \ref{sec:ExptPath} we discuss an experimental-verification path to be followed in order to falsify or, hopefully, verify (!), the above model, thus supporting further the assumption on the potential r\^ole of MT as quantum (bio)computers. Specifically, we discuss the Rabi splitting phenomenon, which is associated with the r\^ole of the entire MT as a cavity, and, if verified, would be a strong indication in favour of the model. We also describe experimental arrangements, and the  pertinent measurements, that would probe quantum coherence and environmental entanglement in individual tubulin dimers, which are the important building blocks of a MT.
For the benefit of the reader we summarise our assumptions on the various parameters of the models, and the pertinent physiological conditions, in two tables. Finally, our conclusions and outlook are given in section \ref{sec:concl}. 

\section{Solitonic effects in Microtubules and observable biological functions}\label{sec:MTqm}

The role of microtubules (MTs) as candidates for coherent energy or signal transduction has been previously proposed in the context of quantum electrodynamics (QED) models~\cite{mn1,nem}. In light of experimental demonstrations of long-lived quantum coherence in light-harvesting complexes ~\cite{collini,algae,EntangledLHCNature2010,QuantumExcitationRev2015}, and inspired by efforts to engineer biomimetic systems such as artificial leaves~\cite{mershinNature2012}, these conjectures warrant renewed theoretical attention~\cite{nmdice10}. While MTs and algal light-harvesting antennae are structurally and functionally distinct, both are intricate protein assemblies. The observation of robust quantum coherence in biological systems at ambient temperatures provides compelling motivation to consider that analogous mechanisms may underlie energy and information transfer in MTs \emph{in vivo}, consistent with the theoretical framework developed in Refs.~\cite{mn1,nem}.

\begin{figure}[ht]
\begin{center}
\includegraphics[width=15cm]{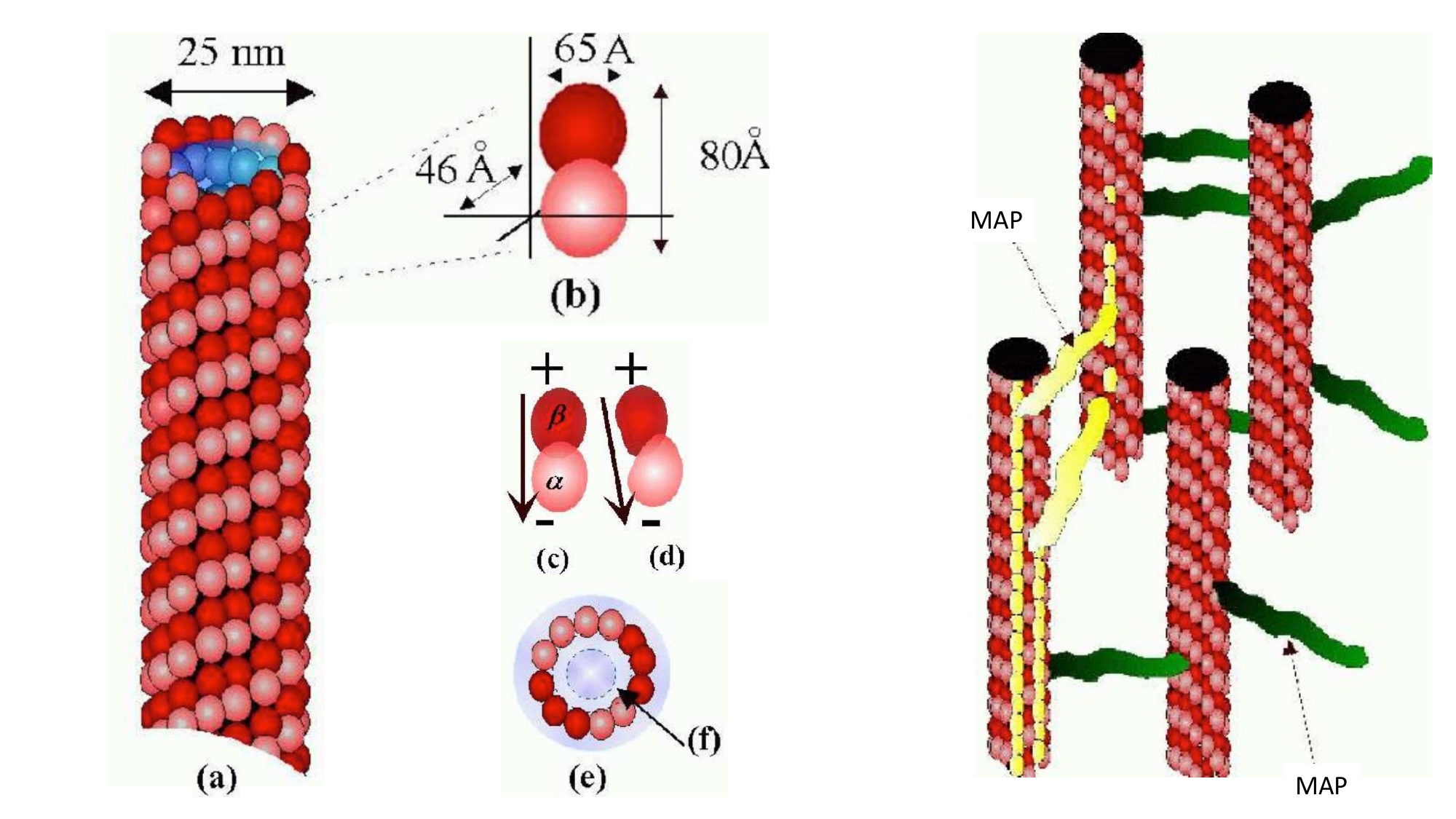}
\end{center}
\caption{\emph{\textbf{Left }}: A microtubule (MT) (a), showing individual dimer subunits and their dimensions (b) (1 Angstrom = 0.1 nm). The walls consist of tubulin protein dimers ((c) GTP tubulin, (d) GDP tubulin), which are arranged usually in 12 or 13 helical protofilaments (vertical chain-like structures, parallel to the long axis of MT). The interior (e) is full of ordered water molecules. In the cavity model of MT!\cite{mn1,mmn1}, a thin interior layer near the dimer walls (f) behaves as a high-Q electromagnetic cavity.
\emph{\textbf{Right Picture:}} A network of microtubules typical of the neuronal cytoskeleton. The ``rungs" cross-connecting  MTs  are microtubule associated proteins (MAP~\cite{MershinFlies2004}) (figures from ref.~\cite{mmn1})}
\label{fig:MT}
\end{figure}
Basic features of these models are briefly reviewed here in a modern context of quantum computation in connection with biological systems~\cite{qc0,qc1,qc2,qc3}.  Microenvironmental conditions,  for \textit{in vitro} or  \emph{in vivo} MT systems, are found to be crucial for the  maintainance of the coherence of quantum effects, when the temperature of the system is within physiological parameters.  \color{black} For the convenience of the reader we give in Table~\ref{tab:glossary} 
a glossary of symbols to be used in what follows, together with their meanings (equation numbers refer to the manuscript). \color{black}

\begin{table*}[t]
\centering
\caption{Glossary of symbols (equation numbers refer to the manuscript).}
\begin{tabular}{lllc}
\hline
Symbol & Meaning & First defined / used & Units \\ \hline
$\varepsilon$ & Relative dielectric constant (MT interior/medium) & Eq.\,\eqref{eow}, Table \ref{tab:bioquantum_parameters} & -- \\
$E_{\mathrm{ow}}$ & r.m.s.\ ordered-water field amplitude & Eq.\, \eqref{eow} & V\,m$^{-1}$ \\
$\omega_0$ & Tubulin dimer transition frequency & Sec.\,\ref{sec:Rabi} Eq.~\eqref{rabi} & s$^{-1}$ \\
$\omega_c$ & Dominant cavity mode frequency & Eq.\,\eqref{homegac} & s$^{-1}$ \\
$\Delta$ & Detuning, $\Delta=\omega_0-\omega_c$ & Eq.\,\eqref{rabi} & s$^{-1}$ \\
$\lambda_0$ & Single-dimer vacuum Rabi coupling, $dE_{\rm ow}/\hbar$ & Eq.\,\eqref{l0} & s$^{-1}$ \\
$N$ & Number of dimers in the cavity mode & Eq.\,\eqref{dimerN} & -- \\
$\Omega_{\pm}$ & Absorption peaks (Rabi branches) & Eq.\,\eqref{rabi} & s$^{-1}$ \\
$t_{\mathrm{decoh}}$ & Decoherence time (cavity-limited) & Eq.\,\eqref{owdecoh} & s \\
$J_{ij}$ & Dipole--dipole coupling & Eq.\,\eqref{Jij} & energy \\
$h,g_0,g_1,g_2$ & Pseudospin $\sigma$-model parameters & Eqs.\, \eqref{contlag}--\eqref{L2} & -- \\
$\kappa$ & $g_2/g_1$ (phase-diagram parameter) & Sec.\,\ref{sec:MTqm} (below Eq.\, \eqref{L2}) & -- \\
$\sigma$ & $(h-g_0)/(2g_1)$ (phase-diagram parameter) & Sec.\,\ref{sec:MTqm} (below Eq.\,\eqref{L2}) & -- \\
$\Sigma_0$ & Area of hexagonal unit cell & Fig.\,\ref{MT1b} & nm$^2$ \\
$\rho$ & Areal mass density (continuum limit) & Eq.\, \eqref{contlag} context & kg\,m$^{-2}$ \\
$u_0,\ \zeta_0$ & Velocity parameters ($\zeta_0^2=1-1/h$) & Eq.\,\eqref{vel} & -- \\
$k$ & Elliptic modulus of snoidal solutions & Eqs.\,\eqref{J1}--\eqref{kink2}, Fig.\,\ref{Sn2a} & -- \\
$\xi$ & Soliton solution argument  & Eq.\,\eqref{xidef} & -- \\
\hline
\end{tabular}
\label{tab:glossary}
\end{table*}

MTs are fundamental constituents of most eukaryotic cells and all neurons~\cite{mt}, playing a crucial in the cell structure, growth, shape and mitosis. They have a cylindrical shape and 
typically consist of 13 (and in some cases 14) protofilaments 
 (see fig.~\ref{fig:MT}). They are formed by the spontaneous polymerization of heterodimers built of two globular proteins (tubulins). The tubulin protein dimers are characterized by two hydrophobic pockets, of length 4 nm = $ 4 \cdot 10^{-9}$~m each (the total length of a dimer being $\sim$8 nm shaped like a peanut in a shell), and they come in two conformations, alpha ($\alpha$) and beta ($\beta$) tubulin, depending on the position of the unpaired electric  charge of 18 e relative to the pockets, which is responsible for the generation of significant electric dipoles. The internal cylindrical region of the MT (which contains  ordered-water~\cite{ordered}) has diameter 15 nm, while the external cross section diameter spans 25 nm. 
MTs can grow up to 50 $\mu m$ long (with an average length of 25 $\mu m$). Each MT is built of a set of macroscopic dipoles which generate dynamical electric fields. The latter prove crucial for an understanding of the functional properties of MTs and their interactions in biological systems. 

In \cite{Sataric1}, the formation of one-spatial dimensional solitons in simplified ferroelectric models of MT has been studied from a rather phenomenological point of view. It has been argued in that work that such solitonic structures, which were assumed  propagating along the MT main symmetry axis, provide efficient energy-transfer mechanisms. These solitons are kinks of an appropriate variable, associated with the appropriate projection (on the main MT axis) of the electric dipole displacement vectors between the two tubulin dimer alpha and beta conformations. 

In a series of works~\cite{mn1,nem}, we have developed a microscopic \emph{quantum electrodynamics cavity model} for MT and proposed phenomenology and experimental pathways towards validation~\cite{towardsTests}.
In our model, we took the full cylindrical structure  into account, together with the ordered-water interior of the MT. A crucial role in our construction is played by the strong dipole-dipole interactions between the ordered-water dipole quanta with the electric-dipole moments of the tubulin dimers. These interactions are strongest for water dipole quanta, near the hydrophobic tubulin dimer walls of the MT, in interior cylindrical regions of about 10 Angstr\"om from the walls. These electromagnetic dipole-dipole interactions are responsible for overcoming thermal losses and are found~\cite{mn1} to be the dominant forces, leading to environmental entanglement and eventual decoherence \`a l\`a Zurek~\cite{zurek} in: 
\begin{align}\label{owdecoh}
t_{\rm ow-decoh} =  {\mathcal O}(10^{-7}-10^{-6})~\rm s\,. 
\end{align}
It is important to stress once again at this point that such a decoherence time is due exclusively to the r\^ole 
of the ordered water in the MT interiors, 
specifically it is assumed in \cite{mn1} that the main source of decoherence 
is the loss of ordered-water dipole quanta through the imperfect MT cavity walls, made out of tubulin dimers.
The decoherence time \eqref{owdecoh}
is much longer than, e.g. the one advocated in the analysis of \cite{Tegmark2000}, 
where the approach of \cite{ph} to consciousness has been criticized), 
of order in the range $t_{\rm decoh~MT~estimate} \in 10^{-20} - 10^{-13}~{\rm s}$, depending on the specific environmental source.  The upper limit of this short decoherence time  has been considered in \cite{Tegmark2000}  as a conservative estimate, 
corresponding to the case in which the main decoherence-source are the Ca$^{2\, +}$ ions in each of the 13 MT protofilaments. Although, for reasons stated, we disagree that such a short decoherence time applies to the QED cavity model of MT~1\cite{mn1,nem},
nonetheless we point out that such short decoherence times are not far from the decoherence times \eqref{algaedecoh}, which proved  
sufficient for the Algae antennae to quantum compute the optimal path for information transduction across distances of order 2.5 nm. As we shall argue below, such short decoherence times might also be sufficient for a `decision making' process on behalf of basic groups of heterodimers in a MT which may constitute the unit of quantum computation (qu(D)it), see discussion below in section \ref{sec:qucom} in such systems.

The basic underlying mechanism for dissipation-free energy and signal transduction along the MT is the formation of appropriate \emph{solitonic} dipole states in the protein dimer walls of the MT, which are reminiscent of the quantum coherent states in the Fr\"ohlich-Davydov approach. These dipoles states are classical, obtained after decoherence of quantum states, and correspond to solutions of the non-linear equations that describe the dynamics of the MT within certain models that take proper account of the dipole-dipole interactions.   
\begin{figure}[ht]
\begin{center}
 	\scalebox{0.5}{\includegraphics{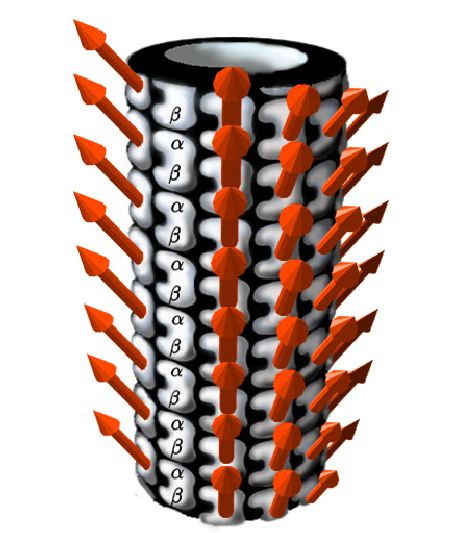}}\\
  \end{center}
 \caption{The structure of the cytoskeleton microtubule (MT).  The arrows indicate the orientation of the permanent dipole moments of the  tubulin heterodimers with respect to the MT surface. The permanent dipole moments of the tubulin dimers are all oriented in such a way that the spherical polar angle of the dipole vectors with respect to the symmetry axis of the MT (assumed here along the $z$ direction) is approximately~\cite{TBCC} $\Theta_0 \simeq 29^{\,\rm o}$. Picture  from ref.~\cite{mexico}.
  \label{MT}}
  \end{figure}
  
In this respect, in ref.~\cite{mexico}, we have discussed the emergence of, and classified, various kinds of solitonic excitations, which arise as solutions of the field equations of appropriate classical field theoretic non-linear systems that model the MT dipole interactions in the tubulin dimer. We have used the full cylindrical geometry of MT, but we did not take into account the complicated ordered-water interior, whose role as mentioned above, was assumed simply to provide -through its dipole interactions with the tubulin dipoles - the long decoherence time \eqref{owdecoh}. So although important in this latter respect, the role of the ordered water was not considered further in \cite{mexico},  where we constructed  a classical non-linear pseudospin $\sigma$-model, which we assumed to describe the coherent (classically behaving) state of the tubulin dipole quanta after the elapse of the decoherence time \eqref{owdecoh} (see fig.~\ref{MT}). 
As discussed in the relevant literature~\cite{TBCC}, in the ground state of an MT,  the orientation of the permanent dipole moments of the tubulin dipoles with respect to the surface of the MT is such that the relative spherical polar angle with the symmetry axis of the MT is $\Theta_0 \approx 29^{\,\rm o}$. 
We suggest here that, during the decoherence time \eqref{owdecoh} (or even in shorter time intervals), in analogy with the situation in Algae, the MT `quantum computes' the most efficient pathway for energy transfer, which is realized by the formation of the appropriate solitons, in the way we describe below.  

As discussed in \cite{mexico}, due to their interaction with the noisy aqueous microenvironment, MTs can experience a strong radial electrostatic field leading to the additional (radial) polarization of tubulins~\cite{BSJ}, as happens, for instance, in the wake of a biomolecular binding event or passage of an action potential. In such cases, it is known that, even inside brain-MT bundles, fields such as those generated by passing action potentials can be felt, and the associated electrical oscillations have been observed experimentally ~\cite{BrainBundleMT})
The total mass of each tubulin heterodimer can be estimated as, ($M \approx 1.84\cdot 10^{-19} \rm g $). Each heterodimer can be considered as effective electric dipole with $\alpha$ and $\beta$ tubulin being as positive and negative side of the dipole, respectively \cite{SM1}.

In the model of~\cite{mexico}, each dipole is treated as a classical pseudo-spin, $\mathbf S_i$, with a constant modulus.  
The lattice model (over the lattice of the dimers) describes the dynamics of tubulin dipoles and their interactions across the two-dimensional MT cylindrical surface. The potential energy of the system can be written as~\cite{mexico}:
\begin{align}
&{U}=S^2\sum_{ \langle  i,j \rangle}    J_{ij}\big( {\mathbf n}_i 
\cdot{\mathbf n}_j  - 3 (\mathbf n_i\cdot {\mathbf e}_{ij} )(\mathbf 
n_j\cdot {\mathbf e}_{ij}) \big)  \nonumber \\
&+ \sum_{i} \big(PS^2 ({\mathbf n}_i \cdot  {\mathbf e}_z	)^2 + QS^4 ({\mathbf 
n}_i\cdot  {\mathbf e}_z	)^4 -
B S \mathbf n_i\cdot {\mathbf e}_r \big).
\label{H2a}
\end{align}
where we have parameterized  the pseudo-spin $\mathbf S_i$ by  the unit vector $\mathbf n_i$, as: $\mathbf S_i = S \mathbf n_i$, where $S$ is the module of $\mathbf S_i$, assumed constant in our approach, as already mentioned. In \cite{mexico}, we took the direction along the $z$ spatial axis to coincide with the main symmetry axis of the MT. 
The quantity $\mathbf e_{ij}$ denotes the unit vector  parallel to the line connecting  the dipoles, the latter being represented by the pseudospin vectors ${\mathbf S}_i $ and ${\mathbf S}_j$. The first term on the right-hand side of \eqref{H2a} describes  the dipole-dipole interaction among the tubulin dimers. The pertinent interaction coupling $J_{ij}$ depends on the inverse cubic power of the distance between dipoles,  according to  the well-known law of electrostatics, 
\begin{align}\label{Jij}
J_{ij} = \frac{1}{4\pi \varepsilon \, \epsilon_0 \, r_{ij}^3} \,,
\end{align}
where $\varepsilon$ is the permittivity of the MT microenvironment in units of that of the vacuum, 
$\epsilon_0$,  and $r_{ij}$ is the distance between sites $i$ and $j$ of the lattice model. As is common, in  \cite{mexico} we assumed that $J_{ij}$, are nonzero only for the nearest-neighbor dipole moments (in practice, next to nearest neighbor dipole-dipole interactions are considered suppressed). The middle term, with $P$ and $Q$ appropriate interaction couplings, has the form of a double-well quartic on-site potential, and takes into account~\cite{THST} the assumed ferroelectric properties at physiological temperature ranges for the MT~\cite{Sataric3,TBCC}, and their effects on the  effective spin, $\mathbf S_i$, while the last term , describes the effects of the transversal (radial) electrostatic field with amplitude $B$ acting on the dipoles, which is produced by the solvent environment of the MT. Thus, all the further effects of the ordered water molecules (apart from their important contribution to lead to the long decoherence time \eqref{owdecoh}), are captured by this term. 

\begin{figure}[tbh]
 \begin{center}
\scalebox{0.17}{\includegraphics{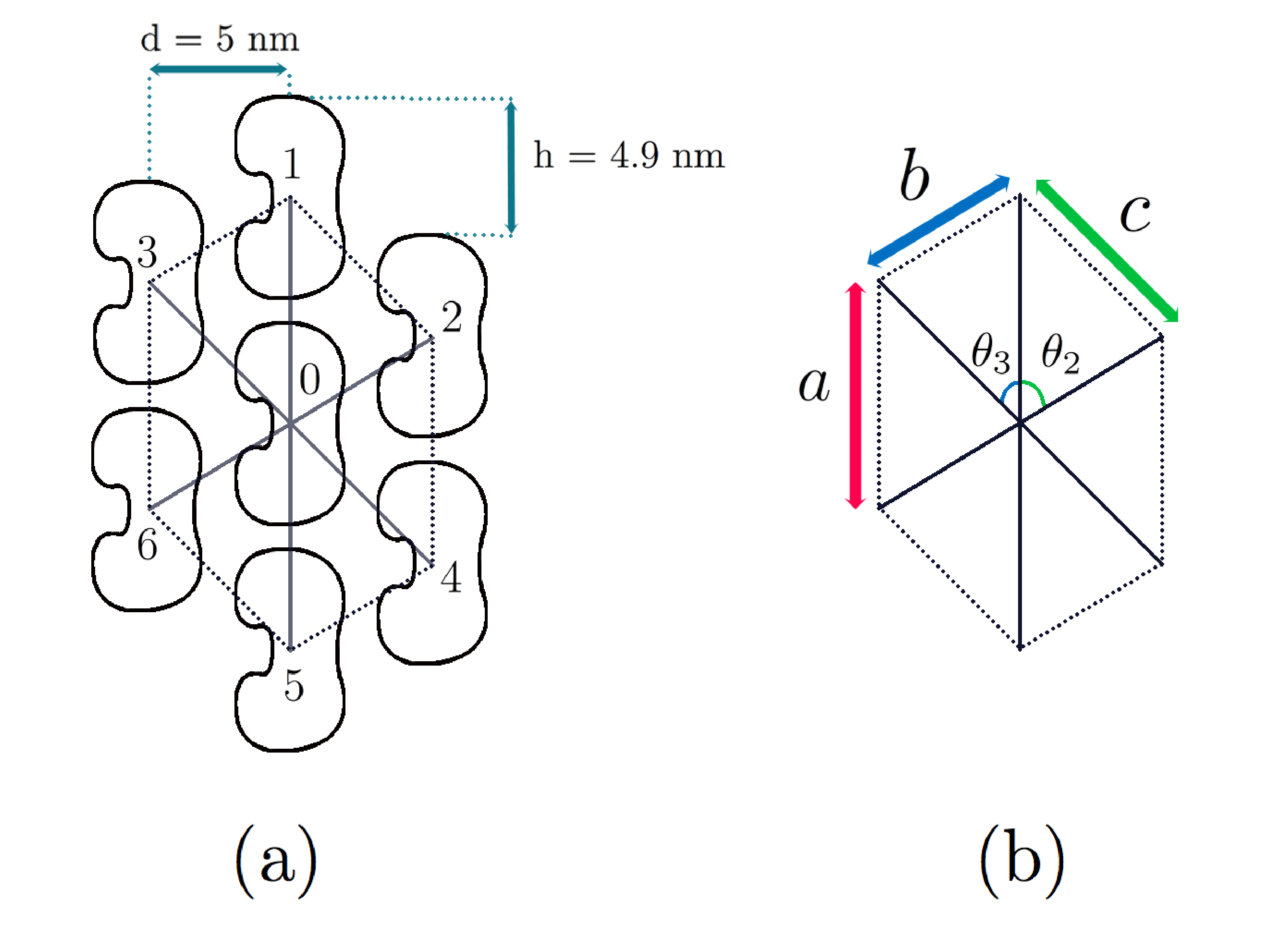}}
 \end{center}
\vspace{-7mm}
 \caption{Tubulin neighborhood in the hexagonal unit cell of the microtubule.  The distance between dimers is $d$. The heterodimer helix direction is defined 
 by the height, $h$. The typical values of parameters are: $a=8\, \rm nm$, $b=5.87\, \rm nm$, $c=7.02\,\rm nm$, $d = 5\, \rm nm$, $h = 4.9\, \rm nm$,  $\theta_1 =0$,  $\theta_2 =58.2^{\,\rm o}$, $\theta_3 = 45.58^{\,\rm o}$ \cite{Sl2,TBCC,THST,GTJT}. This  structure will play the r\^ole of the quDit basic  information storage unit in our modelling of the MT as a biocomputer.}
 \label{MT1b}
  \end{figure}

 The system of MT dimers may be represented as a triangular lattice, as shown in Fig. \ref{MT1b}, so that each spin has six nearest neighbors.   The constants of interaction between the central dipole (labelled ``$0$'')  in Fig. \ref{MT1b} and its nearest neighbors are denoted as $J_{0\alpha}$, and  the distance between the central spin and its nearest neighbors as $d_{\alpha}$ ($\alpha =1,2, \dots, 6$). We set $d_{01}= d_{04} =a$, $d_{02}= d_{05} =b$, $d_{03}= d_{06} =c$. The corresponding angles (between the central dimer and others) are denoted as, $\theta_1$, $\theta_2$ and $\theta_3$, so that: $ {\mathbf e}_{01}\cdot   {\mathbf e}_{01} =\cos\theta_1$, $ {\mathbf e}_{01}\cdot   {\mathbf e}_{02} =\cos\theta_2$, $ {\mathbf e}_{01}\cdot   {\mathbf e}_{06} =\cos\theta_3$. Typical values of parameters known from the literature are: $a=8\, \rm nm$, $b=5.87\, \rm nm$, $c=7.02\,\rm nm$, $\theta_1 =0$,  $\theta_2 =58.2^{\,\rm o}$, $\theta_3 = 45.58^{\,\rm o}$, $S=1714$~Debye~ \cite{Sl2,THST} (See Fig. \ref{MT1b}b.) The radius of the MT can be estimated as, $R \approx 11.2 \,\rm nm$ \cite{TBCC,GTJT}. The unit cell shown in Fig. \ref{MT1b} consists of the central spin surrounded by six neighbors. Its area is: $\Sigma_0= 3ad= 120\, \rm nm^2$.  

As discussed in \cite{mexico}, the continuum approximation proved sufficient for classifying and studying the solitonic solutions arising from the non-linear lagrangian  corresponding to the pseudospin non-linear $\sigma$-model with interaction potential given by \eqref{H2a}. 
Using the local spherical coordinates $(\Theta_i,\Phi_i)$ to define the orientation of the dipole,
$\mathbf n = (\sin\Theta \cos\Phi, \sin\Theta\sin\Phi, \cos\Theta )$,
the continuum Lagrangian of the system reads:
 \begin{align}
	\mathcal L & =  \frac{\rho}{2 } ((\partial_t\Theta)^2 + \sin^2 \Theta 
	(\partial_t\Phi)^2) + \frac{1}{2}\big(\big (\nabla \Theta\big )^2 +\big 
	(\nabla \Phi\big )^2\big) \nonumber \\
	&-\frac{h}{2}(\cos \Theta \sin \Phi\nabla \Theta + \sin \Theta \cos 
	\Phi\nabla \Phi)^2 -\frac{h}{2}\sin^2 \Theta (\nabla \Theta)^2  - 
	\mathcal W(\Theta, \Phi),
	\label{contlag}
	\end{align}
with the interaction potential $\mathcal W(\Theta, \Phi)$ assuming the form~\cite{mexico}
\begin{align}
\mathcal W(\Theta, \Phi) &=   ( g_0 - h)\cos^2\Theta  \nonumber \\
&+ {g_1} \cos^4\Theta 
- h\sin^2\Theta \sin^2\Phi - g_2\sin\Theta \cos\Phi\, ,
\end{align}
with $h= (6S^2/J )\, \sum_{a=1}^3 J_{0 a}\, {\rm cos}^2 \theta_a$, 
$g_0 = P\, S^2/J, \, g_1= Q\, S^4/J, \, g_2= B\, S/J$, and $J= 2\, S^2\, \sum_{a}^3 J_{0 a}$\,. On introducing the dimensionless coordinates, $\zeta = z/\sqrt {\Sigma_0}$ and $\tilde R = R/\sqrt {\Sigma_0}$, the continuum Lagrangian of the system becomes that of an anisotropic 
$\sigma$-model:
\begin{align}
	{\mathcal L} &=\frac{\rho}{2 } \bigg(\frac{{\partial\mathbf n} }{\partial 
	t}\bigg )^2 + \frac{1}{2} (\nabla \mathbf n )^2
	 - \frac{h}{2}(\nabla n^2 \cdot\nabla n^2 + \nabla n^3 \cdot\nabla 
	n^3) - {\mathcal W}(\mathbf n) ,
	\label{L2}
\end{align}
with 
${\mathcal W}(\mathbf n) =  h( n^1)^2 + g_0(n^3)^2  +g_1( n^3)^4 - g_2 n^1$, with the order 
parameter, $\mathbf n$ being  the local polarization unit vector specified by a 
point on the pseudospin sphere, $S^2$.

In \cite{mexico} the zero-temperature phase diagram of this model has been studied in detail, yielding for the ground state (which is characterized by a permanent dipole moment) a paraelectric and a ferroelectric phase, separated by the line $\kappa = 4\sigma $, where $\sigma  \equiv   (h-g_0)/(2g_1)$ and $\kappa \equiv  g_2/g_1$. 
The ferroelectric phase of the MT ground state occurs for $\sigma  >  0$ and $\kappa= g_2/g_1 < 4\, \sigma$, while the paraelectric phase occurs in the regime of parameters $\sigma<0$, and $\kappa >4\sigma$. This zero-temperature paraelectric phase corresponds to the radial orientation of the permanent dipole moments of the tubulin dimers with respect to the surface of the MT.  For finite temperatures of interest to realistic MT systems, we refer the reader to \cite{TBHM} where the critical order-disorder transition temperature depends on the values of the dipole moment and the electric permittivity of the system. It will not be of further impact to our considerations in this paper. We only remark for completion that a discussion on the importance of ferroelectricity in biological systems has been given in \cite{zioutas}, and it has been at the heart of the concrete MT modelling since the early days~\cite{Sataric1}, \cite{mn1}.

\begin{figure}[ht]
  \begin{center}
 \scalebox{0.40}{\includegraphics{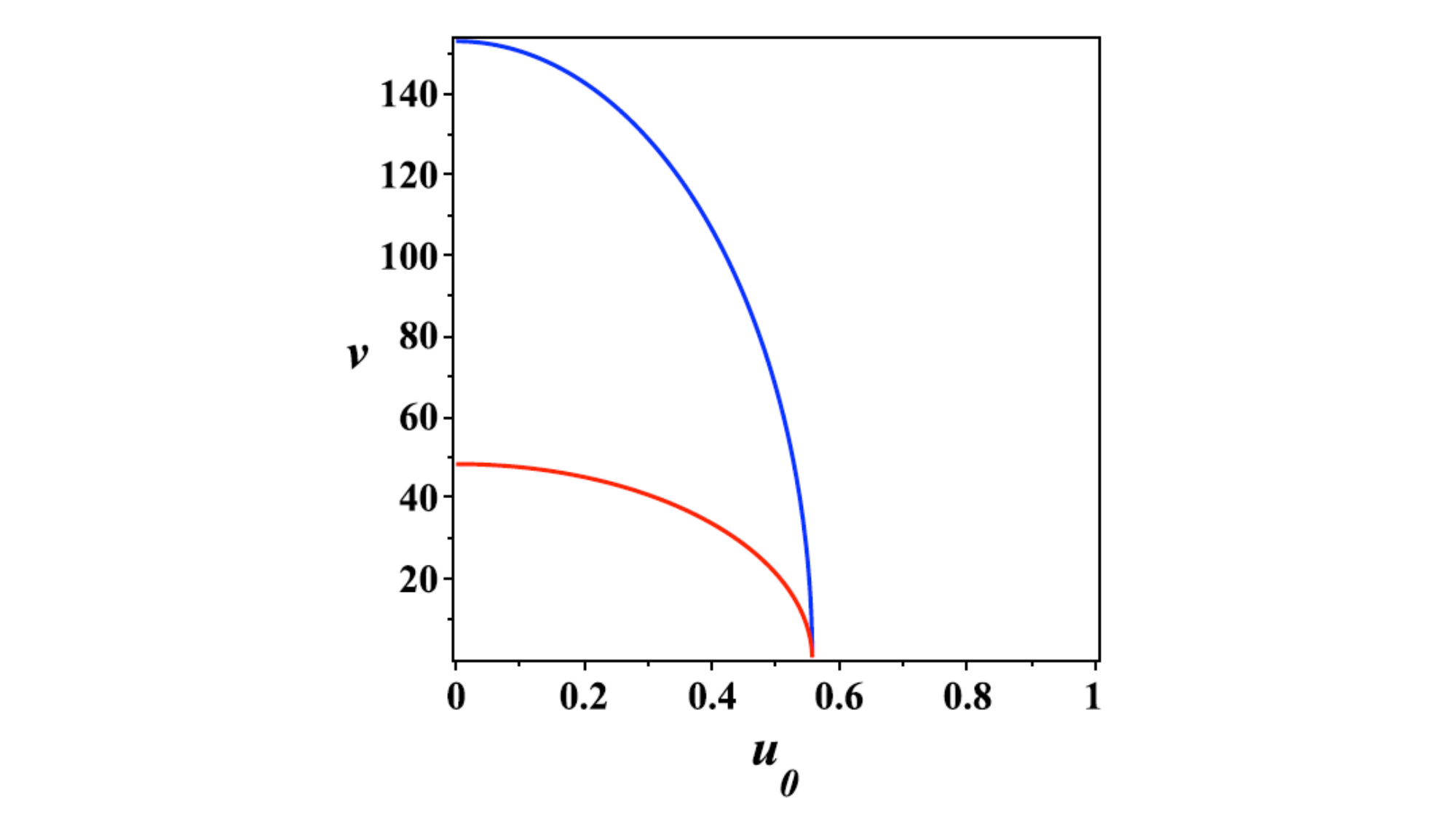}}
\end{center}
 \caption{\color{black} The (bounded) velocity \eqref{vel} (in m/s) of the non-linear wave as a function of the variable $u_0$, for two different values of the tubulin dimer mass (viewed as a dipole of length $l$) $M=10^{-22}\, \rm g$ (red) and $M=10^{-23}\, \rm g$ (blue), with $l \simeq 2$~nm. Picture taken from \cite{mexico}.\color{black}}
 \label{fig:maxvel}
  \end{figure}

The classification of the (finite energy) soliton solutions of the Lagrangian system \eqref{contlag} is of interest, and as discussed in \cite{mexico}, there are kink, snoidal waves, spikes and helicoidal static soliton and also waves propagating along the MT.  Of specific relevance to our case are the helicoidal waves due to their stability, but also due to the general applicability of such helical structure models to many scales and sizes in biology from the alpha helices and chirality of small signaling molecules to the ubiquitous helices of DNA and RNA and their numerous variants in most chiral biopolymers. In order to construct solutions of the equations of motion for nonlinear waves moving along the MT with a constant velocity, $v$, we use the traveling wave ansatz. We assume that in cylindrical coordinates the field variables are functions of  
 \begin{align}\label{xidef}
 	 \xi = \sqrt{\frac{2}{\eta p\Sigma_0}}(z + h_0 \varphi/2\pi -v t),
 \end{align}
 where  $\eta = h/g_1$ and $p=1+(h_0/2\pi R)^2$. On defining $u = \cos \Theta$, one can then show that the field equations possess the first integral of motion~\cite{mexico}: 
	\begin{align}
	& (u_0^2 - \cos^2 \Theta  ) \bigg( \frac{d \Theta}{d \xi} \bigg)^2 + \sin^2 \Theta   \Big (u_0^2 - \frac{1}{h} \cot \Theta -\sin^2 \Phi \Big ) \bigg( \frac{d \Phi}{d \xi} \bigg)^2 \nonumber \\ &+ \frac{1}{2}\sin (2\Theta) \sin (2\Phi) \frac{d \Theta}{d \xi} \frac{d \Phi}{d \xi}  - (\sigma - \cos^2\Theta)^2   +  \eta\sin^2\Theta \sin^2\Phi  + \kappa\sin\Theta \cos\Phi	= \rm const,
	 \label{TW1a}
	  \end{align}
	  where  $u_0^2 =  1- 1/ h-\rho v^2/(hp \Sigma_0)$. This implies  for the nonlinear wave propagation velocity $v$:
\begin{align}\label{vel}
	v=\sqrt{(\sigma_0^2-u_0^2 )\frac{hp\Sigma_0 }{\rho}},
\end{align}
where we set $\sigma_0^2 = 1 -1/h$. 
For completeness, we mention that the analysis of \cite{mexico}, taking into account the parameters entering the model of MT under consideration here, shows that the velocity of the wave is bounded from above  $v \leq v_0$, where $v_0 \approx 155 \rm m/s$ \color{black} (see figure~\ref{fig:maxvel}).\color{black}

There are {\it one-dimensional solutions} characterised by $\Phi=0$, which propagate along the symmetry axis of the MT, in similar fashion to the one-dimensional solitons of the initial simplified models of MT~\cite{Sataric1,mn1,nem}. 
In that case, we choose for convenience the constant of integration on the right-hand-side of eq.~\eqref{TW1a} (upon setting $\Phi=0$) as $\varepsilon =  (\sigma - u_0)^2$. We then observe that the solutions are snoidal waves and kinks, corresponding to the case where $\kappa=0$, which implies the absence of the intrinsic electric field ($g_2 =0$). 

The solution corresponding to a snoidal wave, is given by the following expression:
\begin{align}
u =k \,{\rm sn}(\xi -\xi_0,k).
 \label{J1} 
\end{align}
Here $k= \sqrt{2\sigma -u_0^2}$, and ${\rm sn}(z,k)$, $z \in \mathbb C$, is the Jacobi elliptic function~\cite{abramo}. Hence, the sn waves exist when  $u_0^2<2\sigma < 1+u_0^2$. The period $T$ of the sn-wave is proportional to 
the complete elliptic integral of the first kind \cite{abramo}: 
\begin{align}
T=4\, \int_0^{\pi/2}\frac{d 
	\varphi}{\sqrt{1- k^2 \sin^2 \varphi}}\,.
\end{align}

\begin{figure}[tbh]
  \begin{center}
 \scalebox{0.36}{\includegraphics{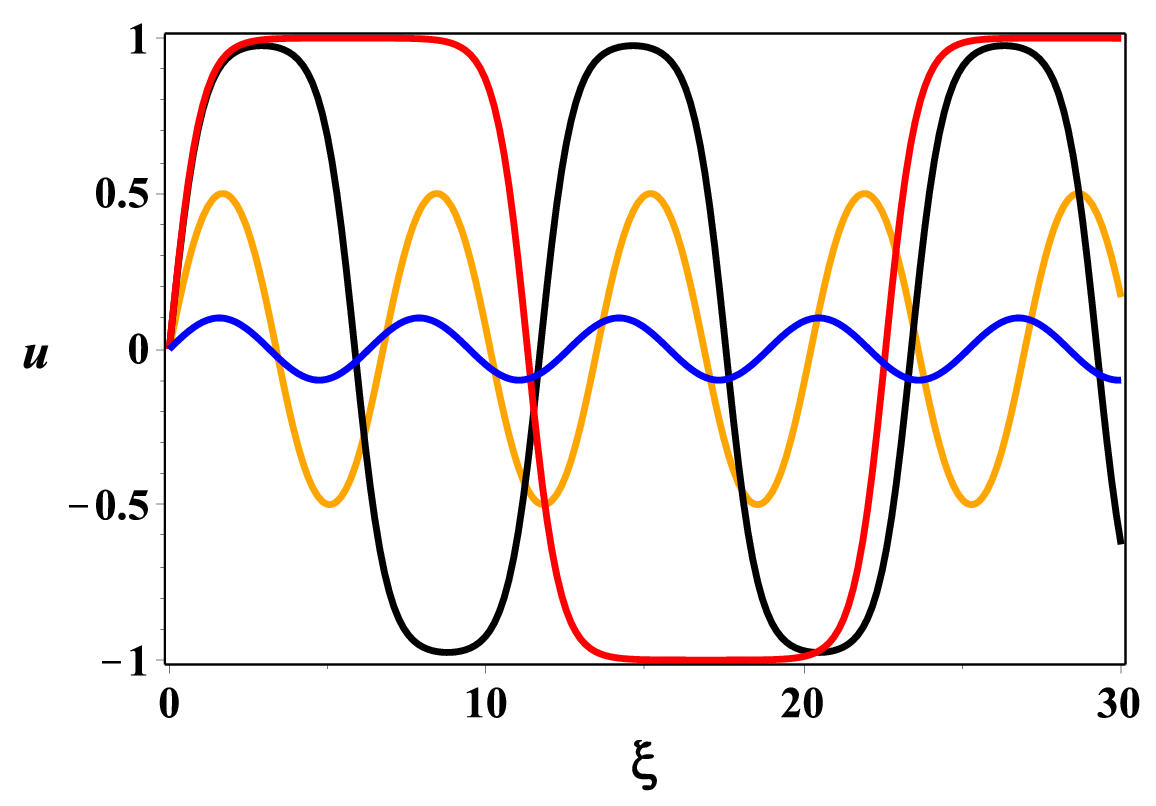}}
\end{center}
 \caption{The  sn-solution: $k = 0.1$ (blue), $k = 0.5$ (orange),  $k = 0.975$ (black), $k = 0.9999$ (red). Picture taken from \cite{mexico}.
 \label{Sn2a}}
  \end{figure}

The static sn-solutions for different choices of the constant $k$ are depicted in Fig.~\ref{Sn2a}~\cite{mexico}.   
For $k^2 \ll 1 $ and $k'^2 = 1- k^2 \ll 1$. It can be seen~~\cite{mexico} that the solutions go to zero smoothly $ u \to 0$, as $k, k^\prime \to 0$, whilst when $k=1$  the sn-waves become the \emph{kink} ({\it cf.} \eqref{kink}):
\begin{align}\label{kink2}
	u = \tanh (\xi - \xi_0),
\end{align}
with the boundary conditions: $u(\pm \infty) =\pm 1$.\footnote{Kinks, as is well known~\cite{solitonsrev,MND}, 
and discussed in the introduction of this article, 
admit a  topological classification in terms of the appropriate homotopy group. In our case, the topological charge, $\pi_0$, of the kink \eqref{kink} is determined by the magnitude, $n_z$ of the polarization vector at the ends of the MT:
\begin{align}
	\pi_0 = \frac{1}{2}(n_z(+\infty) -n_z(-\infty)).
\end{align}
To change the topological charge one needs to overcome the potential barrier, proportional to the size of the MT (formally, infinite potential barrier).}

We remark at this point that such one-dimensional solitons have been considered in connection with dissipation-free  energy and signal transduction in phenomenological one-dimensional models of MT in \cite{Sataric1,mn1}. In \cite{mexico}, such solutions have been derived from realistic three-dimensional lattice models, entailing dipole-dipole interactions on the dimer walls of a MT. 

Apart from kinks, another interesting solution for the case 
$\kappa =0$ are {\it Spikes}. The latter are  excitations of the ground state. Estimates of the energy carried by a spike have been provided in the analysis of \cite{mexico}. 
The electric field produced by the spike can be estimated as~\cite{mexico}, 
$\Delta E_z = E_z^{\rm max} (u_{sp}^2 - u_g^2)^2$, where $ E_z^{\rm max} = \frac{J g_1}{S}$ is the maximum value of the electric field due to the permanent dipoles, which is reached when 
all dipoles are aligned along the MT (in which $u_g =1$).
The maximum value of the electric field produced by spike  $\Delta E_z \leq \Delta E_z^{\max} $, 
has been estimated in \cite{mexico} as 
$\Delta E_z^{\max} = E_z^{\rm max} (1 - u_g^2)^2 = { E_z^{\rm max}} \cos^4\Theta_0 \leq E_z^{\rm max}$,  where $\Theta_0$ denotes the angle between the permanent dipole and axis orthogonal to the surface of the MT.
Notice that the maximum magnitude of the electric field produced by spike is bounded by $E_z^{\rm max}$. 
As discussed in the literature~\cite{TBCC},  and mentioned above, in the ground state the orientation of the dipoles with respect to the surface of the MT are $\Theta_0 \approx 29^{\,\rm o}$. Taking into account  data from \cite{mexico}, one can arrive at the following estimation for the electric field produced by the spike: $\Delta E_z^{\max} \approx  0.6 { E_z^{\rm max}} $. To evaluate $E_z^{\rm max}$, we use  data available for the electric field inside of  the MT: $E_z \sim 10^5 \div 10^8 \,\, \rm V/m$ \cite{Sataric1}. Then, we arrive at the following estimate for the electric field produced by the spike: 
$\Delta E_z^{\rm max} \lesssim  \,0.6\cdot (10^5 \div 10^8 )\,\, \rm V/m ~.$

In addition to solutions with $\Phi=0$, there are also solitonic configurations with $\Theta=\frac{\pi}{2}$, $\Phi \ne 0$, which are chiral solitons that are related to the paraelectric ground state. 
Finally, there are also two-dimensional solutions, with both $\Phi \ne 0, \Theta \ne 0$, which have the form 
$\Theta = \Theta(z+ \nu \varphi - vt)$ and
 $\Phi= \Phi(z+ \nu \varphi - vt)$.  Such solutions describe  two-dimensional nonlinear waves propagating on the MT surface along the $z$-direction.  Among the solutions, are two-dimensional kinks, static helicoidal snoidal solutions, and a {\it helicoidal} sn-wave, which is of central interest to our discussion here. For details we refer the reader to \cite{mexico}. In fact, there could be travelling helicoidal solitonic solutions
 combined to a double helix along a MT, mimicking the structure of DNA molecules, which are known to be particularly stable~\cite{dnastability}. 
 
In our picture, as already stressed, we view the above classical solutions as various outcomes of coherent superposition of quantum states of dipoles. 

\section{Microtubular networks as logic gates}\label{sec:MTgates}

Above we have reviewed  work on soliton solutions arising in the non-linear dynamics of dimer dipoles in microtubular biosystems modelled by pseudo spin non-linear $\sigma$-models. The presence of such solitons, if confirmed by Experiment~\cite{experiments}, would serve as a critical step towards our understanding of energy and signal transduction by (these) biological entities.  
\begin{figure}[tbh]
  \begin{center}
 \hfill \scalebox{0.25}{\includegraphics{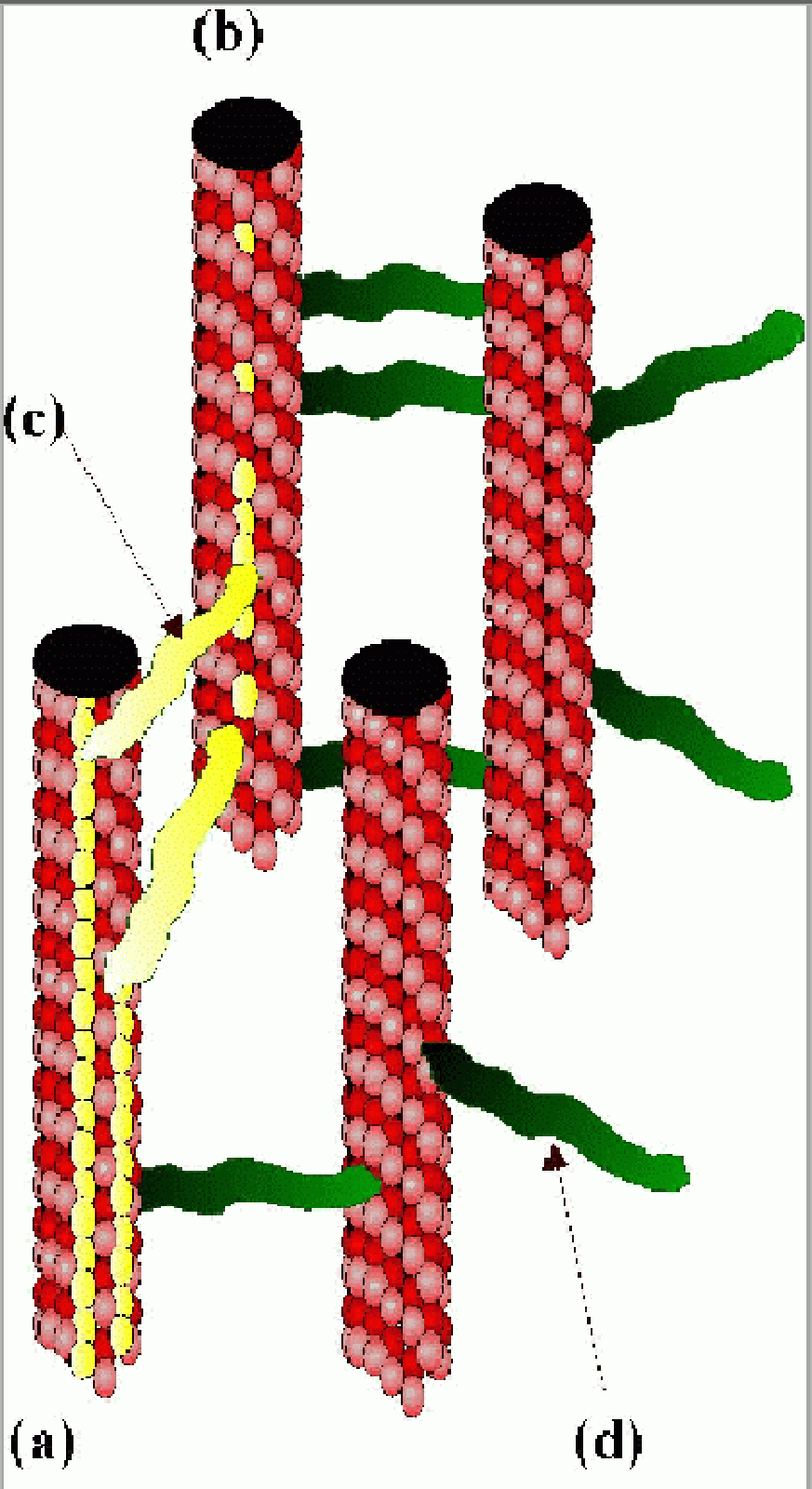}} \hfill \scalebox{0.40}{\includegraphics{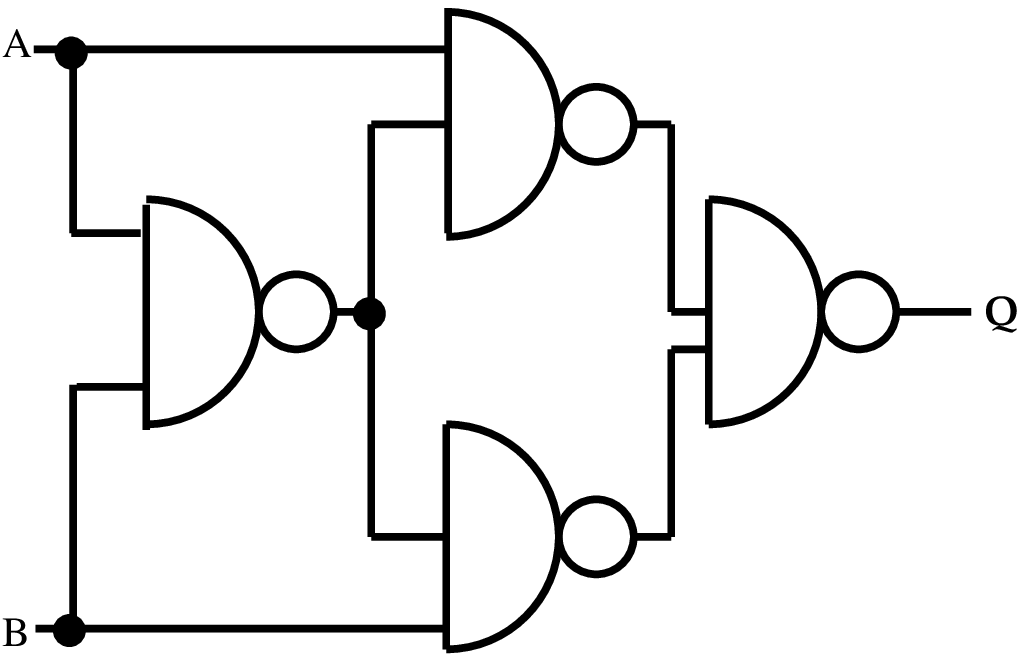}}
 \end{center}
 \caption{A MT arrangement in cell as a `logic' XOR gate. Left panel: the biological arrangement,  a group of MTs and their MAPs, reproduced from ~\cite{mmn1}. \color{black} The MT (a) acts as an ``input'', whilst (b) acts as an ``output'', with (c) denoting a MAP transmitting a soliton, while (d) represents a ``quite'' map. \color{black}
 Right panel: a conventional XOR gate for comparison. \color{black} A and B act as input, while Q is the output. 
 For a XOR, the truth table reads: A=0, B=0, then Q= 0; If any of A,B is 1 and the other 0, then Q=1, while if both A and B are 1, then Q=0.
\color{black}
 \label{gates}}
  \end{figure}
  
In addition to efficient energy transport, the presence of solitonic structures in MT may imply their role as biological logic gates, as proposed initially in~\cite{nem}, and elaborated further, from a quantum computational viewpoint in~\cite{mmn1} (See fig.~\ref{gates}, left panel). Although MT do not themselves branch,  the analogue of a `logic' XOR gate (see Fig.~\ref{gates}, right panel) by MT arrangements in cells has been proposed with  microtubule associated proteins (MAPs) that connect the various MTs in a network given the role of information storage ~\cite{MershinFlies2004, towardsTests, MershinFlies2011}. Once a soliton is formed along one MT, an ``active MAP'' (yellow colour in left panel of Fig.~\ref{gates}) can transport it from one MT to another. In the MT arrangement depicted on the left panel of Fig.~\ref{gates}, an XOR logic gate can be realized provided the  ``0" entry is represented by the absence of a soliton and the ``1" entry by the presence of a soliton. In this arrangement, MT (a) acts as the ``Input'' MT, whilst MT (b) is the ``Output'' MT. (c) is a MAP transmitting a soliton, while (d) represents a ``quiet" MAP (green coloured MAPs). MT (a) has two solitons travelling (yellow colour), encountering two MAPs (yellow coloured MAPs) that transmit both  solitons to MT (b). The solitons would arrive out of phase at MT (b) and cancel each other out.   The truth table for XOR reads: 
$0,0\rightarrow 0; \,    
0,1 \rightarrow 1;  \, 
1,0 \rightarrow 1; \, 
1,1 \rightarrow 0\,,$  
and in this case is realized by MTs if the MAPs are arranged in such a way that each can transmit a soliton independently but if they both transmit, then the two solitons cancel each other out. We stress that the existence of snoidal waves is crucial for the behaviour of MT as logical gates, since for such solitonic structures, the out-of-phase cancellation 1,1 $\, \rightarrow \,$ 0 exists automatically.

\section{Scalable Quantum Computation in Microtubules}\label{sec:qucom}

A scalable quantum system is defined as the one that can grow from a few qubits 
to thousands or even millions, while maintaining its performance before the scaling. Sometimes this may require an integration with some classical computing resources.   In this section we shall attempt to address this question with reference to the Microtubules, which we shall argue can behave like scalable quantum computers, under some circumstances that we shall specify. 
In fact, as we shall argue below, the basic storage information unit of an MT when viewed as a quantum coherent system is a quDit, that is, a higher-dimension qubit~\cite{qc2,qlg}.
To have a complete mapping of any system into a quantum computing device, the following important issues should be addressed~\cite{critique}:
(i) a precise description of the quantum states of the basic information storage unit (qubit or quDit),
(ii) a description of the mechanism through which the wavefunctions representing these states become entangled,
including specification of the basis in which measurements of the qub(D)it’s properties are performed in situ, and
(iii) a means of achieving quantum coherence over the required time scale.
It is the purpose of this section to address such questions and present a consistent model of MT as a quantum biocomputer. 

To this end, we first remark that above we have treated the solitons on a MT as classical solutions, and the XOR gate role of MT arrangements was construed according to classical physics and computation. As in our previously published work ~\cite{mn1,nem}, these solitons are treated as \emph{macroscopic quantum coherent states}, or at most as (not completely decohered, minimum entropy) pointer states~\cite{zurek2}, 
which  survive long enough so that processes such as energy transmission (and by extension memory storage) along MTs of length of a few $\mu$m are allowed to take place. In our approach, this can happen under fine-tuned conditions~\cite{mn1},
that  defend the  system  from environmental decoherence losses. The dipole-dipole interactions between the tubulin dimers and ordered-water dipole quanta~\cite{giudice} in the neighborhood of the (hydrophobic) walls of the  MT tubulin dimers are crucial in accomplishing this. 

\subsection{Proposed mechanisms for biocomputation: dipole-dipole interactions between MT dimers, water microenvironment and external fields }\label{sec:expt2}

At a quantum level, these dipole-dipole interactions of a tubulin dimer with dipole $\vec p$ and its fluid environment can be expressed via interactions of the form \eqref{H2a}, \eqref{Jij}, where the $S_i$ now are viewed as components of a quantum spin (pertaining to the dimer and the ordered-water dipole quantum) and not a pseudospin as in the classical model of \cite{mexico}, discussed  in section \ref{sec:MTqm} above. For concreteness, the relevant interactions between tubulin dimer (TD) and ordered-water dipoles (OW) read schematically:
\begin{align}\label{owmtdip}
U \ni \sum_{i\ne j} \frac{1}{4\pi\, \varepsilon \, \epsilon_0\, r_{ij}^3} \widehat S_i^{\rm TD} \cdot \widehat S_j^{\rm OW}\,,   
\end{align}
where now $\widehat S_{i,j}^{\rm TD, \rm OW}$
are quantum  operators corresponding to spin vectors, and $\varepsilon$ represents an average permittivity (in units of the vacuum permittivity $\epsilon_0$) of the MT microenvironment including the ordered-water interior. In view of its cubic scaling power with the distance between dipoles (the dominant interactions, as expounded upon in \cite{mn1}),  which are strong enough to overcome thermal losses at room temperatures, are those in a region near the dimer walls, of small thickness 20 Angst\"oms. This provides an analogy of the MT as isolated, high-quality QED cavities~\cite{mn1}, and lead to decoherence times of order given in \eqref{owdecoh}. 
The quality factor of the cavity depends crucially on the strength of the interaction \eqref{owmtdip}, which also involve the dependence on the permittivity parameter $\varepsilon$. The smaller the $\varepsilon$, the stronger the dipole-dipole interactions, and thus the longer the decoherence time of the system of tubulin dimers, before its collapse to one of the solitonic states~\cite{mexico} discussed in section \ref{sec:MTqm}, most likely, from the point of view of stability, to the double helix as we shall discuss below.

We remark  at this point that in the {\it ferroelectric phase} of the system of the tubulin dimers~\cite{tus,mn1,mexico}, 
$\varepsilon$ may increase significantly compared to its value in normal non-ferroelectric environments, especially at temperatures near the critical temperature of the pertinent phase transition. In principle this 
would have an effect in shortening the decoherence times, especially in room temperatures, which is supposed to be the order of the critical temperature of the MT systems viewed as ferroelectric/ferrodistortive ones~\cite{tus,mn1,mexico}. 
In the approach of \cite{mn1}, this effect is compensated by the presence of the thin regions near the dimer walls where the ordered-water-tubulin-dimer interactions \eqref{owmtdip} become very strong, due to the $r_{ij}^{-3}$ scaling. 

In a different setting than the QED-Cavity MT model of~\cite{mn1}, the authors of \cite{Xiang}, 
consider the mean-field electromagnetic (Coulombic) interaction between a single-tubulin-dimer dipole moment $\bf p$ and the net electric charges, of density $\rho (\bf r)$, in its aqueous cellular environment as the main source of decoherence. Such an interaction arises from the net (positive and negative) electric charges at the surface of a dimer, which via Debye shielding,  cause the formation of a {\it counterion} layer, 
on top of the dimer-surface charged layer. This counterion layer contains  unpaired (positive and negative) charges and electrically neutral (mainly water) molecules. 
In \cite{Xiang}  it was assumed that the cellular environment behaves as a plasma, characterised by a Debye length $\lambda_D$ and plasma frequency $\omega_p$, 
which describes the collective oscillations of ions. The counterion layer has a thickness of order $\lambda_D$. The plasma-like cellular environment of the MTs contains quantum excitations of the charge density (plasmons).
When these plasmons are in an excited state, the electrical neutrality of the cellular aqueous environment of the tubulin dimers is destroyed, and net charges appear. 

These charges interact electromagnetically (via Coulomb forces) with the dimer dipole, leading  to a contribution to the Hamiltonian of the system of the form:
\begin{align}\label{dimerenv}
    \mathcal H_{\rm dipole-environ} = \int d^3{\textbf r}^\prime \, \rho ({\bf r}) \, \frac{{\bf p} \cdot \bf {\bf r}^\prime }{4\pi \, \varepsilon \, \epsilon_0 \, |{\bf r}^\prime |^3} \,,
\end{align}
where bold face quantities denote three vectors. The authors of \cite{Xiang}, as in \cite{mn1}, assumed that $\varepsilon \, \epsilon_0 \sim 80$ (the dielectric constant of water at room temperature). 
This approach is different also from that found in \cite{Tegmark2000}, where the cellular environment of the dimers of a MT has been unrealistically simulated by a single distant ion. Nonetheless, even in the more detailed approach of \cite{Xiang}, the  decoherence time of a tubulin dimer, resulting from the  interaction \eqref{dimerenv}, lies in the range 
\begin{align}\label{fs}
  t_{\rm dimer-environ} \in (1, 100)~{\rm fs}\,,
\end{align}
which is much shorter than the corresponding decoherence time in the QED-cavity model of \cite{mn1}, \eqref{owdecoh}.  The two-orders-of-magnitude uncertainty in the value of $t_{\rm dimer-environ}$ in \eqref{fs} are attributed to the corresponding uncertainties of various parameters of the model, including  the consideration of $\varepsilon$ in the range (24, 240),
in the phenomenological analysis of \cite{Xiang}. 
We note that the upper limit of such decoherence times are not far from the ones established by experimental observations in Algae~\cite{algae}, \eqref{algaedecoh}.

The analyses of \cite{Xiang} and \cite{Tegmark2000}, ignore the important role of the ordered-water dipole quanta~\cite{giudice}. As we have discussed in \cite{mn1} and mentioned briefly above, in the thin regions  near the dimer walls of the MT,  there are strong dipole-dipole interactions of water dipole quanta with the dipole quanta of the tubulin dimers,\eqref{owmtdip}, which overcome/shield  the interaction \eqref{dimerenv} and  lead to the behaviour of MTs as high-quality QED cavities, resulting in much longer decoherence times \eqref{owdecoh}, since the basic assumption of the cavity model for MT is that environmental decoherence  occurs mainly due to leakage of ordered-water dipole quanta from the MT dimer walls~\cite{mn1} (we discuss briefly some experimental aspects of the model in section \ref{sec:Rabi}).

Following \cite{mn1}, we assume that the temperature $T$ of the system of MTs we consider is in the range of room temperatures, e.g. $T = {\mathcal O}(300)~\rm K$, while the permittivity appearing in \eqref{owmtdip},  $\varepsilon \simeq 80$, that is, of order of the dielectric constant of water.

An information manipulation system based on tubulin dipole quanta as the substrate for bioqubits could also provide the basic substrate for quantum information processing inside a (not exclusively neural) cell or lab-borne microfluidic arrangement. In a typical MT network, there may be of the order of $10^{12}$ tubulin dimers. A question arises as to whether such large aggregates  of ``subunits'' could be quantum entangled, with the entangled state being maintained for a usefully long time. This question has been answered in the affirmative at least in atomic physics, where  experiments~\cite{caes} have demonstrated the existence of long-lived entangled states of {\it macroscopic} populations of Cs gas samples, each sample containing $10^{12}$ atoms, or in liquid-state quantum computing experiments~\cite{bulkspin}, where entanglement among 
even larger populations of appropriate subunits  is generated via interaction with the electromagnetic field at various frequencies. 

\subsection{Microtubules as Quantum Computers: Detailed Scheme}\label{sec:expt3}

One of the basic questions in quantum computation concerns the nature of the basic unit, the qubit or potential higher-(D)-``dimensional'' extensions thereof (quDit~\cite{qc2}). We argue below that such a basic quDit is provided by the basic hexagonal cell ({\it cf.} Fig.~\ref{MT1b}) of the tubulin dimer (distorted honeycomb) lattice in an MT.\footnote{We remark at this point that honeycomb lattices are also the basic structure of the Carbon nanotubes (CNT)~\cite{nanotubes}, which can be thought of as sheets of graphene~\cite{graphene} rolled up 
in a cylinder. These systems have remarkable structural stability, and an extraordinary combination of mechanical, thermal and electrical properties, including superconducting behaviour, which imply their great potential as energy harvesting and storage devices. 
We may draw several analogies between CNT and MT, as far as their geometric characteristics are concerned (we also mention that recently graphene layers have  been used recently in quantum computation~\cite{graphqc}).} 

This is to be contrasted with the initial quantum picture of MT's in \cite{ph}, in which one views the tubulin dimer conformations as providing quantum states $|0>$ or $|1>$, corresponding to the  $\alpha$ or $\beta$ conformations~\cite{ph}. The latter are viewed as {\it identical} for all the dimers, hence the macroscopic coherence in such a system originates from the {\it orchestrated reduction}~\cite{ph} of the collective wave-functions of the ``identical'' dimers, each viewed as a two-state quantum system. 
However, upon taking into account the environment of the unpaired charges in each tubulin dimer, and the associated physiological, as well as geometrical, differences among the dimers, as indicated by the various parameters in fig.~\ref{MT1b}(b), one is tempted to assign internal degrees of freedom to the various dimers of the fundamental hexagonal cell. In this viewpoint,  therefore, the appropriate formalism to describe the basic unit for storage information, and hence for quantum computation in an MT, would not be a binary qubit system, but a higher-dimension quDit~\cite{qc2,qlg}. The number of the independent quantum states  actually included in a primary cell such as that depicted in fig.~\ref{MT1b} depends on whether there is a symmetry under reflection (\textit{i.e.} rotation of the cell by an angle $\pi$). 

In the depiction of the dipoles in the present work, we only consider as differences the geometrical characteristic of the individual dipoles, due to the difference in the angles $\theta_1$, $\theta_2$. Thus, the upper three nearest-neighbour states, 
labeled as  1,2,3 in fig.~\ref{MT1b}(a), are treated as identical to the lower ones, 4,5,6. Hence, the independent quantum basis quDit dimer states are the ones at the four vertices of the parallelogram enclosed by the sides $b$ and $c$, which are
labeled as 0,1,2,3, referring to the corresponding dimers in fig.~\ref{MT1b}(a). 

Each of these states is a quantum coherent superposition of $\alpha$ and $\beta$ conformations, but they are distinct due to their ``internal'' degrees of freedom. Our (pseudo)spin model~\cite{mexico} is ideal to provide the fundamental unit for storage of information in case one views the MT network as a quantum computer. The reader should recall at this point that QuDit and spin systems go hand in hand~\cite{quDitspin}.

Therefore, in each fundamental parallelogram of the fundamental cell of fig.~\ref{MT1b}(b), say 0312, where the 
vertices $i=0,1,2,3$ label the nearest neighbours of tubulin dimers depicted in fig.~\ref{MT1b}(a), one encounters a quDit comprised of 
four-qubit entangled states~\cite{qu4it} (since each tubulin-dimer quantum state, before its collapse,  
can be in a superposition of an $\alpha$ and a $\beta$ conformations). 
The four-qubit entangled states of tubulin dimers 
constitute a convenient basis for the description of the various entangled quDit states in our MT Lattice model
(such a basis of four quibts consists of 2$^4$=16 Quantum states):
\begin{align}\label{4qubits}
 |\psi^{\rm 4~\rm qubits}\rangle & = 
a_0 |\alpha\alpha\alpha\alpha\rangle + a_1 \,|\alpha\alpha\alpha\beta\rangle + a_2 \,|\alpha\alpha\beta \alpha \rangle + a_3 \, |\alpha\beta \alpha \alpha \rangle + a_4 \,|\beta \alpha \alpha \alpha \rangle \nonumber \\ &+ a_{12} \,|\alpha \alpha \beta \beta \rangle + a_{13}\, |\alpha\beta \alpha \beta \rangle | + a_{14}\, |\beta \alpha \alpha \beta \rangle + a_{23}\, |\alpha\beta \beta \alpha \rangle | + a_{24}\, |\beta \alpha \beta \alpha \rangle  + a_{34}\, |\beta \beta  \alpha \alpha \rangle  \nonumber \\
& + 
a_{123}\, |\alpha\beta \beta \beta \rangle | + a_{124}\, |\beta \alpha \beta \beta \rangle
+ a_{134}\, |\beta \beta \alpha \beta \rangle + a_{234}\, |\beta \beta \beta \alpha \rangle
+ a_{1234}\, |\beta \beta \beta \beta \rangle\,,
\end{align}
where the coefficients are in general complex numbers, with the constraint that they lead to appropriate normalization of the state $|\psi^{\rm 4~\rm qubits}\rangle$. 
As discussed in \cite{qu4it}, such entangled four-qubit states can be used to provide the area of the two-dimensional fundamental parallelogram, which can be constructed from the specific entangled state 
\begin{align}\label{Astate}
\mathcal A \rangle = |\alpha \beta \alpha \beta \rangle - |\beta \alpha \beta \alpha \rangle\,.
\end{align}
corresponding to 
a specific quantum circuit, described explicitly in \cite{qu4it}. 

In the realistic MT case, 
where the environment of the tubulin heterodimers is taken into account, 
the two 
parallelograms of the fundamental MT lattice cell may be inequivalent, as we have mentioned above. In such a case, one has a more complicated quDit structure, since the fundamental storage of information unit now comprises of the two parallelograms  0321 and 0456 (see fig.~\ref{MT1b}). We leave for future works a detailed exploration of the full potential of the honeycomb MT lattice architecture and the corresponding MT networks, discussed here, for quantum (bio)computation.\footnote{We cannot resist in pointing out, at this stage, that the r\^ole of the hexagonal MT Lattice fundamental unit in providing a sort of coding for the function of MTs as information storage and processing  devices has been pointed out in \cite{koruga}, as discussed by Nanopoulos in \cite{nano}. However,  as argued in the current paper, it is the entangled states of the two parallelograms of the fundamental hexagonal cell ({\it cf.} fig.~\ref{MT1b}) that play a crucial r\^ole as fundamental information-storage and ``decision-making'' units, for efficient information and signal transduction across the MT. This type of entanglement 
also leads to the formation of the double helical solitonic structures of quantum dipoles, which after collapse become double-helix-like snoidal solitonic waves~\cite{mexico}, that are mechanically stable, as in the DNA case~\cite{dnastability}.} 

These quantum states are quantum entangled (``wired dissipationlessly")  during the time interval \eqref{owdecoh} (measured from the moment of the action of an external stimulus to the MT), in much the same way as the bilin molecules of the algae antennae. In algae, such  a coherent wiring / yoking / quantum entanglement occurs over distances of 25 nm, and the decoherence time is a few hundreds  fs.  This is evidently sufficient time for the algae molecule to quantum-compute the optimal path for signal transmission over distances covering half of the extent of the algae light harvesting antenna.  In the case of the  MT, the distances, over which entanglement is expected to survive, are not restricted only to the fundamental cell depicted in fig.~\ref{MT1b}, but, over the entire MT of lengths of $\mathcal O(\mu \rm m)$. This feature is a consequence of the strong dipole-dipole interactions between nearest tubulin neighbours, and also the ordered-water dipole quanta~\cite{giudice,mn1} and the tubulin dimers themselves, which provide a stronger isolation than in the case of algae, thus leading to much longer coherence times \eqref{owdecoh}. Indeed, the reader should notice that, if the snoidal waves propagate with a velocity of order at most 155 m/s, as discussed after \eqref{vel}, then signals are transduced across a micron ($\mu$m)-long MT in times $\mathcal O(10^{-8})$~s, which lies comfortably in the aforementioned decoherence time interval of the QED-cavity model of MT~\cite{mn1}.

This allows for quantum computation of the system of dimer dipole quanta in an MT, involving a ``decision making process" for determining the optimal path for signal transduction along the MT. The decoherence time \eqref{owdecoh} is long enough to allow all these processes to take place in the following order: 
\begin{itemize}
  \item{{\bf (i)}} The \underline{{\it initial entanglement}}:  the system of tubulin dimers in the fundamental hexagonal unit of the (distorted) honeycomb lattice  in an MT gets entangled upon the action of an external stimulus. At this stage we should remark that there is a `democracy' among the fundamental hexagonal units of an MT dimer Lattice. Any unit in the lattice that gets excited by an external stimulus, behaves in the same way, getting entangled (``quantum wired'') with the rest of the units across the MT. 

 \item{{\bf (ii)}} The \underline{{\it decision-making}} process : this implies the choice of the optimal path for energy and signal/or information transduction in a dissipation-loss, efficient way.
 To put it differently, upon the action of an external stimulus, the MT dimers system collapses, within the decoherence time \eqref{owdecoh}, to one of the solitonic states mentioned in section \ref{sec:MTqm}~\cite{mexico}. We stress that we view the ``classical'' solitons as either coherent quantum states, or minimum entropy pointer  states~\cite{zurek2}, which are different from the coherent states, and are associated with incomplete collapse processes. 
 The precise form of solitons  obtained from the relevant collapse process depends on the external stimulus and environmental conditions,
 the kind of process/``computation'' executed by the MT system, 
 as well as the stability of the soliton. The most efficient scheme, of {\it maximal stability} to transport energy and information
 are double helices of left-right moving helicoidal sn-oidal waves, mimicking the stable structures of the DNA~\cite{dnastability},
 as discussed in section \ref{sec:MTqm} (see also footnote~\ref{foot3}). 
In terms of the fundamental quDit of fig.~\ref{MT1b}(a), 
a double helix is formed by the initial entanglement and subsequent collapse of, say, the 2,0,6 dimers, and dimers along this direction, in such a way that a ``left'moving'' helicoidal sn-wave involving those is formed. 
The other branch of the double helix involves the 3,0,4 (and collinear dimers) along the ``right-moving'' helical sn-oidal wave. The double helix extends along the entire MT, while the loss-free energy- or signal transduction takes place within the time interval \eqref{owdecoh}. 
The kinks or sn-oidal waves along the principal axis of the MT, which involve  dimers in the direction of 1,0,5 in fig.~\ref{MT1b}(a), are not as mechanically stable as the double helix, and in this sense they are not as efficient as the former  for the process of energy and signal/information transduction. On the other hand, the localized spike solutions may be relevant for 
memory switching, as discussed in \cite{spikes}. 

\item{{\bf (iii)}} The \underline{{\it energy or information transfer process}}: This takes place during, or even after, the collapse along the double helix soliton, which is a solution of the classical (after the collapse) pseudospin model, proposed in \cite{mexico}, and reviewed in section \ref{sec:MTqm}.
 
\end{itemize}

If the above mechanism is realized in nature, then the tubulin dimer system of (brain) MT, or even the MT networks (see fig.~\ref{fig:MT}), can operate as room temperature biological quantum biocomputers, as far as certain processes in the brain are concerned. We should stress that, given the enormous number, of order of $10^{12}$, of tubulin dimers in a typical MT,  this would imply an enormous computing power on behalf of the MT their networks. We stress once again that for these considerations to be valid, relatively long decoherence times of order \eqref{owdecoh} are required, which characterize the cavity QED model of MT~\cite{mn1,nem}, as a result of the strong dipole-dipole interactions bwtween tubulin dimer dipole quanta and ordered-water dipole quanta~\cite{giudice}. Unfortunately, short decoherence time or order of a few hundreds of fs, 
as those characterizing the models of decoherence of tubulin dimers discussed in \cite{Tegmark2000} or \cite{Xiang}, although appropriate for the light harvesting antennae of Marine Algae~\cite{collini,algae}, where the entanglement extends over distances of order of 40 Angstr\"oms, they are not suitable for quantum computation in MT, where much longer distances of entanglement are required for efficient quantum computation.

We also stress here that our model for quantum computation involved decoherence mechanisms induced by ordinary non-gravitational environments, as expected to be the case for MTs. In this respect we differ in our conclusions from those of \cite{ph}, where quantum gravity is argued to be the main source of decoherence, due to the fact that some critical mass has been reached by the network of tubulin dimers in an MT. In our scenario of cavity QED model of MT~\cite{mn1,nem,mmn1}, the electromagnetic in origin dipole-dipole interactions, are the ones that dominate over any other interaction, including the weak quantum gravitational one, and are responsible for sufficient environmental shielding of the MT so as to guarantee the relatively long quantum coherence times. 

Another question is whether, via dissipation-free information and energy  transmission, quantum computation could be sustained. Here is where the analogy with the "quantum wiring"  encountered in marine algae, as discussed in the introduction~\cite{collini,algae} serves as an appropriate perspective. There is a direct analogy of the MT system with the algae system, provided we account for the effects of ordered water, which, while traditionally underappreciated has  strong enough  dipole-dipole interactions~\cite{mn1},  to provide the necessary environmental isolation so that quantum effects in tubulin dimers experience decoherence times \eqref{owdecoh}, of order $\mu$s -clearly potentially relevant to many other processes taking place in a living cell  as well.

\section{Towards Experimental Verification}\label{sec:ExptPath}

In this section we discuss an experimental path to be followed in order to falsify the aboved theoretical (QED cavity) models of MT~\cite{mn1,nem,mmn1}. By verifying experimentally the most important features of these models, we shall also strengthen the assumption on their potential r\^ole in (quantum) Biocomputation. 

\subsection{Testing the Cavity-MT Model: Rabi-Splitting}\label{sec:Rabi}

One of the first tests of the QED-cavity model of MT, proposed in \cite{mn1} could be the search for the well-established Rabi-splitting phenomenon~\cite{rabi}, which is characteristic of electromagnetic cavities~\cite{haroche}. According to this effect, upon the action of an external ({\it e.g.} laser) field, of frequency $\Omega$ on {\it near-resonant} cavities, containing (quantum) atoms of characteristic frequency $\omega_0$, in interaction with the coherent photon modes inside the cavity of frequency $\omega  = \omega_0 - \Delta$, with $\Delta/\omega_0 \ll 1$,  
the absorption spectrum of the atoms will peak at two frequencies:
\begin{align}\label{rabi}
  \Omega_{\pm} = \omega_0 -\frac{\Delta}{2} \pm \frac{1}{2} \, \Big(\Delta^2  + 4\, \lambda^2 \,\mathcal N \Big)^{1/2}\, , 
\end{align}
where $\lambda$ is the so-called Rabi coupling, pertaining to the atom-photon interactions, viewing the atoms as spin-1/2 two-state quantum systems, and $\rm N$ is the number of atoms inside the cavity. 

\color{black} 
Representing each tubulin in a MT as a two-state ($\alpha$, $\beta$ conformations) ``atom" in a cavity near the dimer walls, induced by the strong dipole-dipole interactions between the dimers and the ordered-water molecules, we covered in detail in \cite{mn1} showing that a Rabi-like phenomenon similar to that in \eqref{rabi} could be used to characterize these entities at ambient temperatures, during the short decoherence time \eqref{owdecoh}, which is itself linked to the magnitude of the Rabi coupling between the tubulin dimers and the coherent modes of the water-dipole quanta~\cite{giudice} which play the role of "photons" in this formulation ~\cite{mn1}. 

Estimating the parameters that enter the computation of the decoherence time in the cavity model of MT is challenging as it depends on a detailed description of the ordered-water molecules. In \cite{mn1} it is argued that the corresponding Rabi coupling $\lambda$ is provided by the expression
\begin{align}\label{l0}
    \lambda_0 \sim \frac{d_{\rm dimer} \cdot E_{\rm ow}}{\hbar}
\end{align}
where $d_{\rm dimer}$ represents the matrix element of the electric dipole of a single dimer, associated with the transition from the $\alpha$ to the $\beta$ conformations (for relevant physical parameter values see Table \ref{tab:parameters}), which provide the binary nature of the dimer quantum state in our framework, as discussed above. The quantity $E_{\rm ow}$ represents a typical root-mean-square (r.m.s.) value of the amplitude of a coherent dipole field mode in the ordered water~\cite{giudice}. As a crude estimate,  taking into account that, in contrast to the dimers which lie in the surface of a MT, the ordered-water dipole quanta exist in the entire water-interior of the MT, we may borrow relevant formulae from quantum optics of dielectric cavities, to represent 
\begin{align}\label{eow}
E_{\rm ow} \sim \Big(\frac{2\pi \, \hbar \, \omega_c}{\varepsilon\, \epsilon_0 \, V}\Big)^{1/2}\,,
\end{align}
where $V$ is the (MT cylindrical) cavity volume, $\varepsilon$ is the dielectric constant of the medium, taken in \cite{mn1} to be that of water, at room temperatures, $\varepsilon \sim 80$ ({\it cf.} \eqref{dimerenv}), and  $\omega_c$ is the frequency of the coherent electromagnetic mode (``photon'' in quantum optics), which here is replaced by the coherent-dipole-quantum mode of the ordered water. As an estimate, in \cite{mn1} we adopted the ``superradiance'' model of \cite{jibu}, in which $\omega_c$ is calculated from the energy difference, $\Delta E_{\rm ow}^{\rm principal}$, between the two principal energy eigenstates of the water molecule, assuming that these are the dominant ``coherent cavity electromagnetic modes'' in this case:
\begin{align}\label{homegac}
   \hbar \, \omega_c \sim \Delta E_{\rm ow}^{\rm principal} \sim 4~\rm meV \, \qquad \Rightarrow \qquad \omega_c \sim 6 \times 10^{12} ~\rm Hz\,.
\end{align}
As discussed in \cite{mn1} this frequency is in the range of the upper bound of the  assumed range of frequencies of quantum oscillations of the tubulin dimers viewed as two-state quantum systems ($\alpha$, $\beta$ conformations), before collapse, as per the original analysis of \cite{ph}, and in \cite{jibu}. Our study of MT as QED cavity models in \cite{mn1} assumed this upper bound, and thus, in this case, the dominant cavity mode and the dimer system were almost in resonance, with a detuning satisfying $\Delta \equiv \omega_0 - \omega_c \ll \omega_0$, where $\omega_0 $ is the frequency of the dimer conformational quantum oscillations. In \cite{mn1} when estimating the relevant number of dimers (``atoms'')
in the cavity, we restricted ourselves to one-dimensional solitons formed along the protofilament of an MT. 

\begin{figure}[tbh]
  \begin{center}
 \scalebox{0.70}{\includegraphics{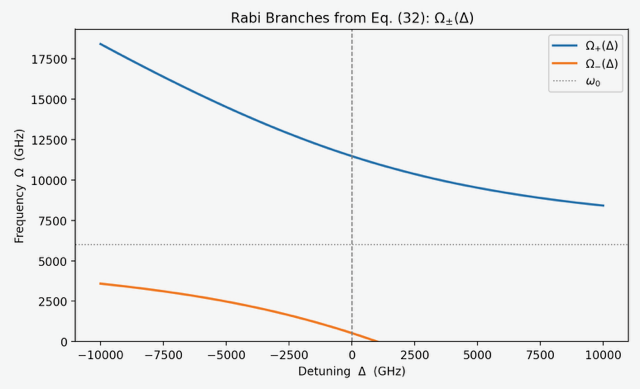}}
\end{center}
 \caption{\color{black} The two (Rabi) branches of the frequency $\Omega$ in \eqref{rabi}, as a function of the detuning $\Delta$ for a 25\,$\mu$m long microtubule: $\Omega_{\pm}=\omega_0-\frac{\Delta}{2} \pm~ \tfrac12\sqrt{\Delta^2+4\mathcal N\, \lambda^2}$ with $\lambda $ the Rabi coupling (Eqs.~\eqref{l0}--\eqref{eow}). Near resonance, the vacuum splitting approaches $2\lambda\sqrt{\mathcal N}$. 
 Physically we require  $\Omega_{\pm} >0$, which restricts appropriately the range of the detuning $\Delta$. 
 \color{black}}
 \label{fig:OmegaDelta}
  \end{figure}

\begin{figure}[tbh]
  \begin{center}
 \scalebox{0.50}{\includegraphics{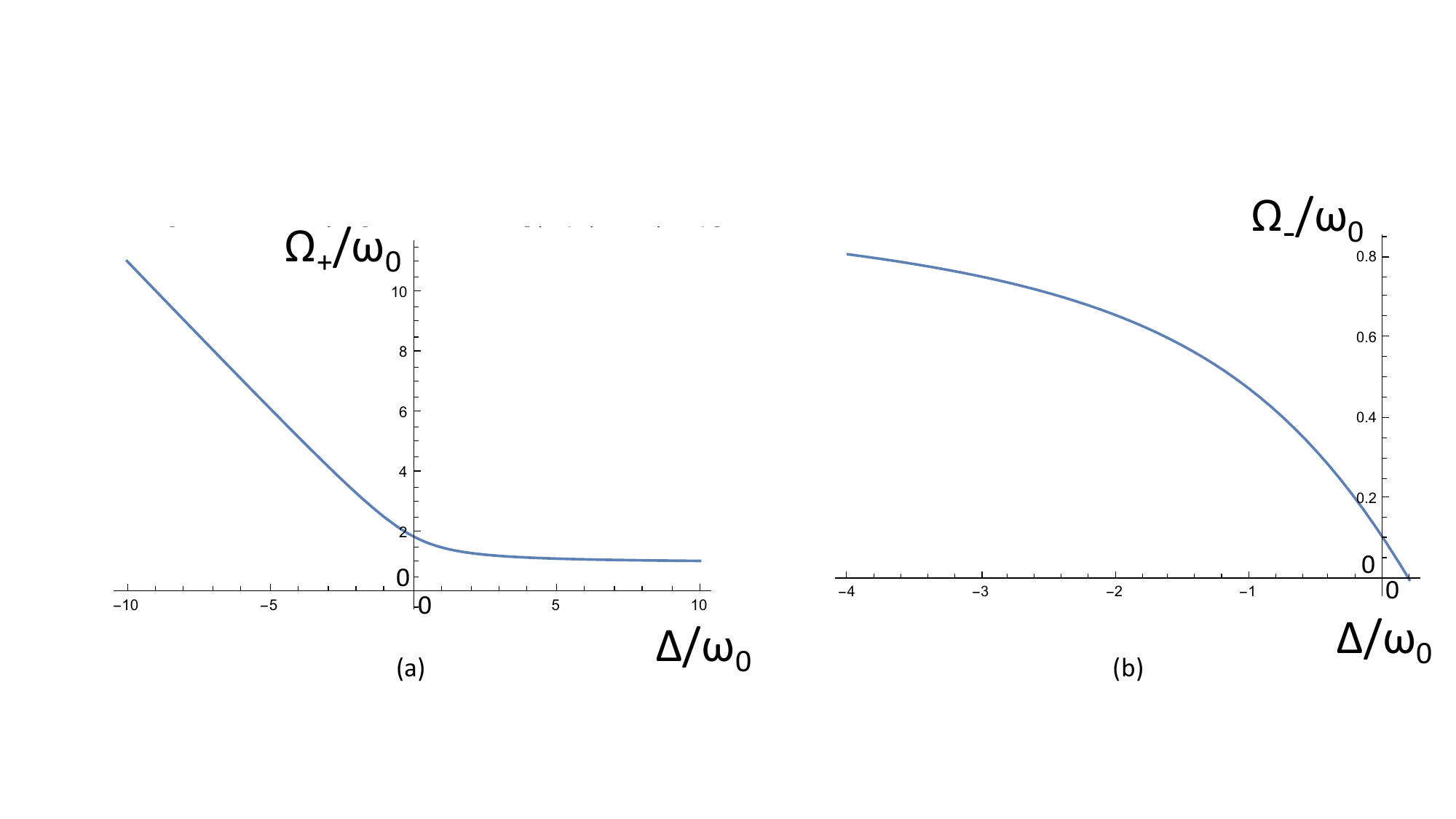}}
\end{center}
 \caption{\color{black} As in figure \ref{fig:OmegaDelta}, the two (Rabi) branches of the frequency $\Omega_{\pm}/\omega_0$ in \eqref{rabi}, as functions of the detuning $\Delta/\omega_0$ for a 25\,$\mu$m long microtubule, 
 with the oscillation frequency of the dimers $\omega_0$ in the $10^{12}$~Hz region,
 but for a bigger range of $\Delta$, to demonstrate the behaviour for large detunings in a clearer way: 
 $\frac{\Omega_{\pm}}{\omega_0}= 1 -\frac{\Delta}{2\omega_0} \pm~ \tfrac12\sqrt{\frac{\Delta^2}{\omega_0^2}+4\mathcal N\, \frac{\lambda^2}{\omega_0^2}}$ with $\lambda^2 \mathcal N^2/\omega_0^2 \sim 0.8$. As in fig.~\ref{fig:OmegaDelta}, the physically allowed range of $\Delta$ is such that the $\Omega_{\pm} > 0$. Left figure (a), the function $\Omega_{+}/\omega_0$. Right figure (b), the function $\Omega_{-}/\omega_0$.
 \color{black}}
 \label{fig:OmegaDelta2}
  \end{figure}

\color{black} In the current work, we consider helicoidal solitons on the MT surface~\cite{mexico}, and therefore we need to consider the number of dimers in the entire MT. For a typical moderately long MT, of length of order of a micron ($\mu$m), consisting of 13 protofilaments, 
there are
\begin{align}\label{dimerN1}
\mathcal N = {\mathcal O}(10^2)\,, 
\end{align}
dimers, each of about 8~nm long. 
For such a MT we have from \eqref{eow}:
\begin{align}\label{eowval}
E_{\rm ow} \sim 1.3 \times 10^{9} \, \Big(\frac{{\rm eV}}{{\rm m}^3\, \epsilon_0}\Big)^{1/2} \sim 17.5\times 10^4\, \frac{\rm V}{\rm m}\,,
\end{align}
where we took into account that the vacuum permittivity is
$\epsilon_0 = 55.26\, e^2 \, {\rm eV}^{-1}\, (\mu {\rm m})^{-1}$. On the other hand, for an MT of length 25 $\mu \rm m$, which we shall make use of in the subsequent analysis for concreteness, 
we obtain 
\begin{align}\label{dimerN}
\mathcal N \sim 3 \times 10^3\,, 
\end{align}
and 
\begin{align}\label{eow2}
E_{\rm ow} \sim 3.6 \times 10^4 \, \frac{\rm V}{\rm m}.
\end{align}

Taking into account that each dimer has a mobile charge (positive or negative) 36$e$, with $e$ the positron charge, with electric dipole moment ~\cite{mmn1}: $d_{\rm dimer} \sim 3 \times 10^{-18}$ Cb $\times$ Angstr\"om. This implies a Rabi splitting for the individual dimers \eqref{l0} of order 
\begin{align}\label{totalRS}
    \lambda_0 \sim 1.0 \times 10^{11}~\rm Hz\,,
\end{align}
and a total Rabi splitting for the entire MT, consisting of 13 protofilaments, of order~\cite{rabi} 
\begin{align}\label{lMT}
    \lambda_{\rm MT} = \sqrt{\mathcal N}\, \lambda_0 \sim 5.5 \times 10^{12}~\rm Hz\,.
    \end{align}
    
\color{black} Moreover, the detuning $\Delta$ is a phenomenological parameter in the current approach, which goes beyond the chain approximation for the MT model presented in \cite{mn1}, where only the case of a resonant cavity 
was considered. The two branches of the function $\Omega(\Delta)$ \eqref{rabi} are plotted in fig.~\ref{fig:OmegaDelta} (and, for a bigger range of values of $\Delta$ in fig.~\ref{fig:OmegaDelta2}), which expresses the Rabi slitting as a function of a general detuning parameter $\Delta$. Such plots will hopefully prove useful in guiding the relevant  experimental searches. The allowed range of $\Delta$ is determined by the requirement of positive absorption frequencies $\Omega_{\pm} > 0$, which is depicted in the figures. 

We appreciate here that an absorption spectrum of MTs in solution or on a surface would be expected. In practice, intact microtubules (MTs) in suspension do not yield a unique, time-invariant UV–visible “absorption spectrum.” Proteins—including tubulin and the same protein when polymerized into MTs exhibit a broad near-UV maximum near 280\,nm arising from aromatic residues (Trp/Tyr) \cite{Agilent2025,Hyvonen2017,AitkenLearmonth}. This intrinsic peak position is not expected to shift with filament length per se; yet upon polymerization  the sample’s strong elastic light scattering (“turbidity”), generally swamps weak absorbance features and is proportional mainly to the mass of polymerized material rather than the exact length distribution \cite{Gaskin1974,HallMinton2005}. Moreover, the hallmark \emph{dynamic instability} of MTs continuously alters the polymer mass and length distribution on experimental timescales, further causing baseline drift and inhomogeneous broadening that preclude a stable spectrum for freely suspended MTs \cite{MitchisonKirschner1984}. For these reasons, the field’s standard quantitative readouts in bulk are turbidimetric (350\,nm) rather than true absorbance spectra \cite{Gaskin1974}, and optical/electrical characterization that avoids turbidity typically relies on well-defined tubulin solutions (e.g., refractometry/SPR for dielectric properties) or immobilized/stabilized MT formats \cite{Biosystems2004}. We therefore have cited representative spectra and parameters for tubulin in the literature that apply in solution (which report the underlying chromophores) and point to stabilized/immobilized MT measurements or dielectric/impedance spectroscopy for polymer-state properties that are not confounded by scattering \cite{Santelices2017}. 

We further note that biochemical heterogeneity among tubulin isotypes and post-translational modifications can slightly alter aromatic content and extinction, reinforcing that reported absorbance amplitudes depend on composition rather than MT length \emph{per se} \citep{Luduena1998}. Indirect Experimental support for microsecond-scale relaxation windows relevant to THz absorption/Rabi splitting in MTs exists.
Direct observation of a vacuum–Rabi doublet in the THz ($=10^{12}$~Hz) regime requires that the driven dipolar polarization of the microtubule (MT)–water system persist for at least a few optical cycles (ps), while practical detectability improves as environmental relaxation slows. Notably, \emph{experimental} AC–impedance measurements on MT \emph{ensembles in electrolyte} show a sharp conductance feature centered at $f\!\approx\!10^{5}$--$3\times10^{5}$\,Hz (FWHM $\sim$\,5$\times 10^{5}$\,Hz), implying a characteristic polarization/solvation relaxation time $\tau\!\sim\!(2\pi f)^{-1}$ in the \emph{microsecond} range for the counter-ion/cloud degrees of freedom that screen tubulin dipoles.\,\cite{Santelices2017} This microsecond window does not claim long-lived quantum superpositions; rather, it shows that the \emph{electrodynamic} channels that would dephase a cavity-coupled MT polarization are not inevitably ps–ns in aqueous media, but can be slowed to $\tau\sim\mu$s by the MT–electrolyte composite. \textit{In vivo} photopharmacology further supports slow dissipation of MT-state perturbations: single 405\,nm pulses halt EB3-tracked polymerization within seconds and recovery occurs on $\sim$\,10--600\,s depending on geometry and model organism, while no Z$\to$E thermal back-relaxation is seen over hours in buffer~\,\cite{Gao2022}. 

Although these cell/animal results reflect biochemical binding and diffusion rather than coherence, they empirically demonstrate that MT-related degrees of freedom can retain an optically imprinted state for times $\gg\mu$s, consistent with the assumption that, under THz drive in a confined EM environment, dephasing relevant to a resolvable Rabi doublet can be bounded from below by microsecond-scale polarization relaxation observed in MT electrolytes. Moreover, THz/near-field optical studies on proteins reveal underdamped, collective low-frequency vibrational modes and long-range elastic motions in the condensed phase, establishing that biomolecular solids support coherent THz excitations when environmental coupling is favorable~\cite{Markelz2014,2DTHz2018}.\footnote{\color{black} For completeness, we mention that large-scale, classical molecular dynamics simulations of MT and their water environment~\cite{simul}, which lead to a theoretical computation of MT absorption spectra in the THz regime, do not seem to show any evidence of large-scale coherent excitations of MT. However, such simulations do not take into account the potential quantum-cavity nature of MT, advocated in \cite{mn1,mmn1}. It is the latter structure that leads to the excitation of quantum coherent states across the MT, upon the action of an external stimulus, and which is held responsible for the potential r\^ole of MTs as quantum biocomputers. Simulating the quantum cavity regions in the ordered-water-filled interior of an MT is not a trivial task, and certainly cannot be done via classical molecular dynamics simulations as those in \cite{simul}.\color{black}}

Some of our results and parameters assumed are tabulated in Tables~\ref{tab:benchmarks_sectionV} and \ref{tab:parameters}.

\begin{table}[h!]
\centering
\begin{tabular}{|l|p{10cm}|}
\hline
\textbf{Quantity} & \textbf{Estimate / Formula} \\
\hline
\textbf{Fields \& coupling} 
& For a 25\,$\mu$m microtubule (inner radius $\sim$7.5\,nm) and dielectric constant $\varepsilon = 80$:
Eq.~\eqref{eow} gives $E_{\rm ow} \approx 3.6 \times 10^{4}$\,V/m; 
Eq.~\eqref{l0} yields $\lambda_0 \approx 1.0 \times 10^{11}$\,s$^{-1}$. \\
\hline

\textbf{Population} 
& $\mathcal N \approx 13 \times L / (8\,\mathrm{nm}) \approx 4.1 \times 10^4$ for $L = 25\,\mu$m (from Eq.~\eqref{dimerN}). \\
\hline

\textbf{Resonant split} 
& $\Omega_{\rm split} \approx 2\lambda_0\sqrt{N}$ (limiting form of Eq.~\eqref{rabi}), which lies in the GHz--THz range for SPR-like fields $10^4$--$10^5$\,V/m. \\
\hline

\textbf{Decoherence window} 
& $t_{\rm decoh} \sim 10^{-6}$\,s (from Eq.~\eqref{owdecoh}), with dielectric sensitivity shown in Fig.~\ref{fig:tdecohepsilon}. \\
\hline
\end{tabular}
\caption{\color{black} Quantitative benchmarks for MTs referenced in Section \ref{sec:ExptPath}.}\color{black}
\label{tab:benchmarks_sectionV}
\end{table}

\begin{table}[h]
\caption{Physical parameters relevant to quantum effects in biological systems. In this table we collect all physical parameter values salient to our calculations and assumptions used in prior sections throughout the present text.}
    \centering
    \begin{tabular}{|c|c|c|}
    \hline
    \textbf{Parameter} & \textbf{Value} & \textbf{Significance} \\
    Tubulin dimer dipole moment & $\sim1.7 \times 10^3$ Debye & Strong electric fields in microtubules~\cite{Biosystems2004}. \\ \hline
    Microtubule diameter & 25 nm & Structural scale of neuronal MT
    \\
    \hline
    Internal microtubule field & $10^5$–$10^7$ V/m & Comparable to semiconductor devices~\cite{Pokorny} \\\hline
    Quantum coherence time  & $10^{-5}$–$10^{-4}$ s & Sufficient to underlie  critical biological processes \\ & & (e.g. cryptochromes in avian magnetoreception \cite{gauger2011}) \\\hline
    Photosynthetic coherence time & $300$ fs & Room-temperature quantum transport \\\hline
    Decoherence time  -    
    MT model of \cite{mn1,mmn1} &
    $10^{-6}$ s    & quantum biocomputation   \\ \hline
    \end{tabular}
    \label{tab:parameters}
\end{table}

We remark at this point that, in typical Rabi splitting situations in quantum optics it is assumed that practically no energy exchange takes place between atoms and cavity modes. In the MT case, this may be guaranteed from the fact that, since the dominant interactions between ordered-water coherent dipole models (``cavity'' modes in this analogue) and dimer-dipole quanta (``atoms'') attenuate with the cubic power of the distance between then, the only dominant interactions are near the walls of the MT, thereby implying that the bulk of the cavity modes (viewing the entire MT as an isolated cavity) does not exchange significant amount of energy with the dimers. This lead us in \cite{mn1} to assume that the main reason of decoherence is the leakage of dipole quanta from the MT interior to the environment, which lead to the principal estimate of decoherence time \eqref{owdecoh}. \color{black} In terms of the microscopic parameters of the model defined above, the decoherence time reads~\cite{mn1}:
\begin{align}\label{qedcavdecoh}
        t_{\rm ow-decoh} = \frac{T_r}{2 n {\cal N}{\rm sin}^2\left(\frac{{\cal N}
n \lambda_0^2t}{\Delta}\right)}\,, 
\end{align}
where $n$ is the average number of oscillation quanta in a coherent mode of dipole, taken in \cite{mn1} to lie in the range  $n = \mathcal O(1-10)$.
The time $t$ appearing in (\ref{qedcavdecoh})
represents the `time' of interaction of the dimer system with the 
dipole quanta, which in \cite{mn1} has been taken to be the average 
life-time of 
a coherent dipole-quantum state. In the 
super-radiance model for the ordered water of \cite{jibu}, which is used in \cite{mn1} to arrive at 
\eqref{qedcavdecoh},
this is 
estimated as
\be
     t \sim \frac{c\hbar ^2V}{4\pi d_{ej}^2 \,\Delta E_{\rm ow}^{\rm principal} \, N_{w}\,L}
\label{lifetime}
\ee
with $d_{ej}$ the electric dipole moment of a water molecule, $L$ the length of the MT, and $N_w$ the number of water molecules in the volume $V$ of the MT. We remind the reader at this point that the quantity 
$\Delta E_{\rm ow}^{\rm principal}$
denotes the energy difference between the two principal energy
eigenstates of the water molecule, which are assumed in \cite{jibu} and \cite{mn1} to play the dominant r\^ole in
the interaction with the (quantized) electromagnetic radiation field. For typical values of the parameters
for moderately long MT, $L \sim  10^{-6}~{\rm m}$, $N_w \sim 10^8$, 
a typical 
value of $t$ is: $t \sim 10^{-4}~{\rm sec} $. 

In \eqref{qedcavdecoh}, we took into account 
that in the QED-cavity model for MT~\cite{mn1},  
the dominant (dimer)-(dipole quanta) coupling 
occurs for ordered-water `cavity' modes which are {\it almost at resonance}
with the dimer oscillators
slightly detuned by $\Delta:~\lambda_0/\Delta 
<<1 $. Moreover, in \cite{mn1} we assumed that the time scale $T_r$ over which a cavity MT dissipates its energy is of similar order as $t$:
\begin{align}
T_r \sim t\,.
\end{align}

Under these approximations, the final expression for the decoherence time \eqref{qedcavdecoh} is:
\begin{align}\label{qedcavdecoh2}
t_{\rm ow-decoh} \sim \frac{2\pi\,\Delta^2\, d_{ej}^2 \, \Delta E_{\rm ow}^{\rm principal} \, N_w \, L}{n^3\, \mathcal N^3\, \lambda_0^4 \, c \, \hbar^2 \, V } = \frac{\Delta^2\, d_{ej}^2 \, \Delta E_{\rm ow}^{\rm principal} \, N_w \, L\, V }{ 2\pi\, n^3\, \mathcal N^3\, d_{\rm dimer}^4 \, \omega_c^2  \, c \, }\,  (\epsilon_0\,\varepsilon)^2  \equiv \mathcal A \, \varepsilon^2\,,
\end{align}
where, in arriving at the last equality on the right-hand side, we made use of \eqref{l0} and \eqref{eow}. 

\begin{figure}[ht]
  \begin{center}
 \scalebox{0.50}{\includegraphics{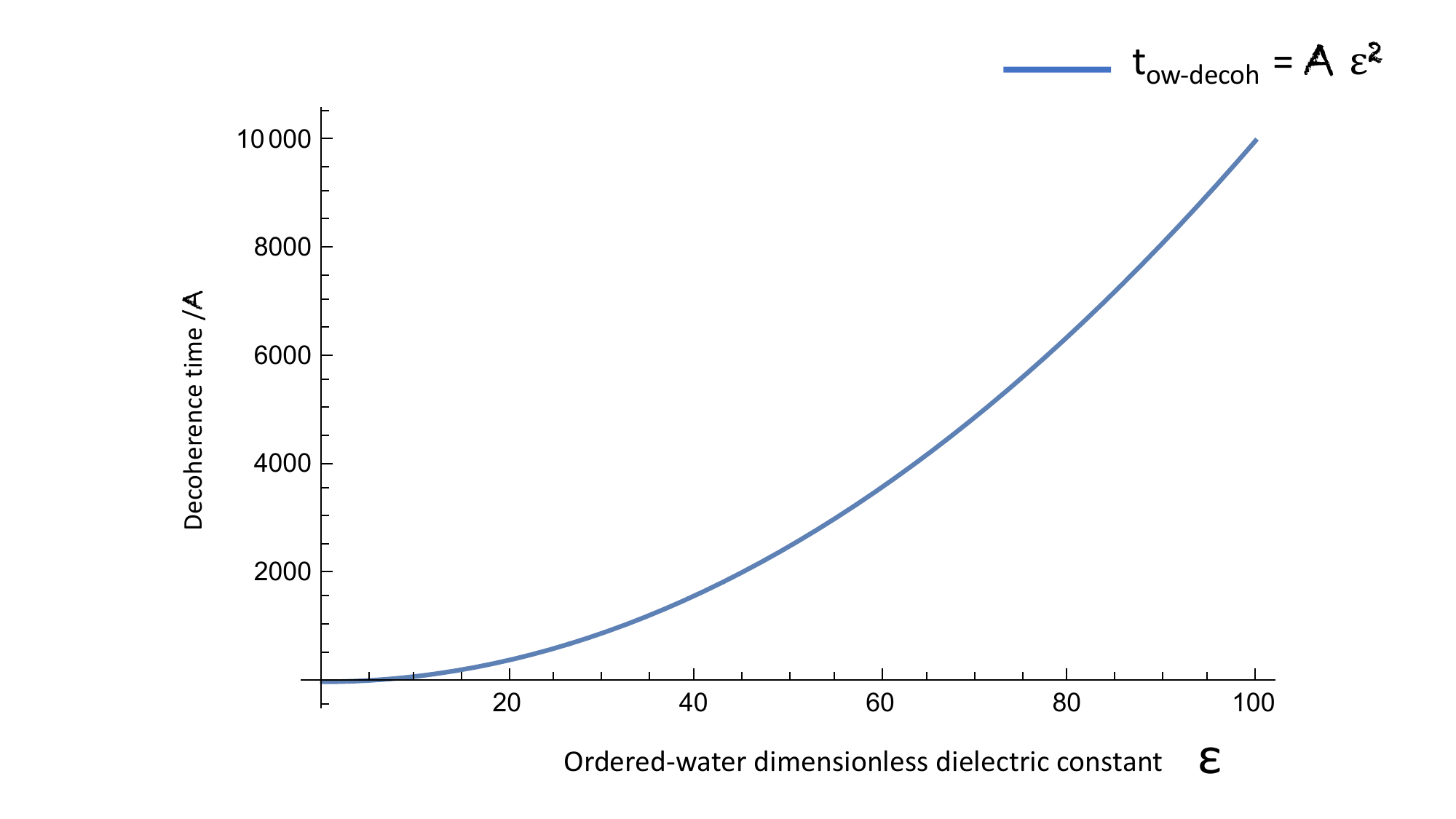}}
\end{center}
 \caption{\color{black} The decoherence time \eqref{qedcavdecoh2} in the QED-Cavity model of MT of \cite{mn1}, as a function of the (dimensionless) dielectric constant $\varepsilon$ of the ordered-water medium. The increase of the decoherence time with increasing dielectric constant $\varepsilon$ is easily understood by the fact that an increase in $\varepsilon$ implies a reduction in the strength of the Rabi interaction coupling \eqref{l0} between dimers and water dipole quanta, which in turn implies a weaker coupling of the system with its environment, and thus a reduction in  the losses through the dimer walls. For the parameters of the model in \cite{mn1}, for $\varepsilon = 80$ the resulting decoherence time is estimated to lie in the range \eqref{owdecoh}, {\it i.e.} $t_{\rm ow-decoh} = {\mathcal O}(10^{-7}-10^{-6})$~s. We also note, for completeness, that there is a slight dependence of  the dielectric constant of water $\varepsilon_w$ with temperature T:
$\varepsilon_w \simeq 80$ at T=20~$^{\rm o}$C (Celcius),  $\varepsilon_w \simeq 73.151$ at T=40~$^{\rm o}$C  and $\varepsilon_w \simeq 55.72$ at T=100~$^{\rm o}$C. The reader should always have in mind, though, that in {\it in vivo} biological systems, which may not always be in thermal equilibrium, the concept of temperature might be subtle, so plots like this in the figure are mainly used to guide in vitro situations.  \color{black}}.
 \label{fig:tdecohepsilon}
  \end{figure}

Upon substituting values for the various parameters of typical MT encountered in biological systems (see Table~\ref{tab:benchmarks_sectionV}) we arrive at the estimate \eqref{owdecoh} for the decoherence time in our QED-cavity model for MTs~\cite{mn1}. This time is much larger than the decoherence time \eqref{fs} of \cite{Xiang}, based on individual dimer dipole-environment interactions, outside the cavity model. Moreover, as can be seen from \eqref{qedcavdecoh2} the decoherence time scales (increases) with the square of the water-environment dielectric constant $\varepsilon \epsilon_0$ (see figure~\ref{fig:tdecohepsilon}),as expected from the fact that an increasing $\varepsilon$ is associated with a decreasing Rabi coupling \eqref{l0}, and hence, a weaker system/environment coupling, implying lesser losses of dipole quanta through dimer walls. The $\varepsilon^2$ scaling in our model is in contrast with that of the model of \cite{Xiang}, in which the decoherence time scales (increases) with the square root of the corresponding environment dielectric constant $\varepsilon$, $\sqrt{\epsilon_0\, \varepsilon}$. The decoherence time \eqref{qedcavdecoh} is also much larger than the decoherence time of the model of \cite{Tegmark2000}, which is of similar order to the one in \cite{Xiang}, given that the main environment of the MT dimer quanta, inducing decoherence, in that work, and also in \cite{Xiang}, is assumed to be the Ca$^{2\, +}$ ions.\footnote{\color{black} Note that in arriving at this scaling law we assume that the electric dipole moments of the dimers and the water molecules do not have a scaling law with the corresponding dielectric constants. Indeed, it is the strong dipole-dipole interactions between the water molecules, due to their bent shape and unequal sharing of electrons that are responsible, combined with hydrogen bonding, for the large value of the corresponding dielectric constant $\varepsilon =80$. \color{black}} Par contrast, in our cavity-MT model, it is the strong isolation of the dimer quanta from other environmental entanglement, except the loss of coherent dipole-quanta modes through the imperfect walls of the MT, that leads to such long decoherence times \eqref{owdecoh}. The reader is reminded that this isolation is provided by the strong diopole-dipole interactions between the dimers and the ordered-water molecules. 
\color{black}

\subsection{Probing quantum coherence and environmental entanglement of individual tubulin dimers}\label{sec:entdimers}

We assess the feasibility of experimental quantum information processing in neuronal microtubules by modeling tubulin heterodimers as multilevel quantum systems (quDits) capable of sustaining entanglement. By comparing the coupling strengths of tubulin dipoles to both evanescent surface plasmon fields and transient electric fields from neuronal action potentials, we find interaction energies within experimentally accessible regimes. These results support the use of photonic probes, such as surface plasmon resonance (SPR), for detecting coherent dipole dynamics in microtubules anchored to functionalized surfaces. Given their intrinsic dipole moment and quasi-periodic cylindrical geometry, stabilized microtubules—particularly those organized by microtubule-associated proteins (MAPs) in axonal architectures—emerge as structurally viable candidates for biologically compatible quantum information substrates \cite{mmn1,mn1,mexico}. As derived earlier, solitonic excitations such as snoidal or helicoidal waves that can propagate along MTs emerging from coherent dipole alignments \cite{mexico}.

A framework for quantum information processing arises when treating each tubulin dimer as a multi-level quantum system---a quDit---with its state space shaped by conformational, electrostatic, and spatial degrees of freedom. Unlike standard qubits, which encode information in binary $|0\rangle$ and $|1\rangle$ states, these quDits may operate across $D > 2$ discrete states due to the geometrical asymmetry and interaction potentials in each MT unit cell \cite{mn1,qlg}. This assignment is justified by the distinct roles played by the seven (including the central one, see fig.~\ref{MT1b})  heterodimers discussed previously, comprising the hexagonal MT lattice unit, which differ in angular orientation, dipole alignment, and microenvironmental exposure \cite{mexico}.

Dipole--dipole interactions between adjacent tubulin dimers, modeled as quantum spin operators, form the effective mechanism for quDit coupling. These interactions are short-range and scale with $r^{-3}$, favoring nearest-neighbor interactions and enabling controlled state entanglement within a unit cell and across adjacent MT filaments \cite{mn1,mexico}. The presence of ordered water molecules within the MT lumen further enhances environmental shielding and extends decoherence timescales to the microsecond regime under physiological conditions \cite{mn1,giudice}, potentially allowing biologically useful quantum operations to occur before classical collapse.

\begin{table*}[h!]
\centering
\caption{Compiled physical, chemical, and quantum parameters relevant to the study of microtubules as potential substrates for quantum information processing. All values correspond to physiological ('ambient") conditions we have defined as ($37^{o}\,{\rm C}$, pH 7.2, salinity $\sim 150$~mM), and where applicable, values are normalized or expressed in units of thermal energy ($kT$), GTP hydrolysis energy, or standard quantum energies. experimentally) Dielectric constant of tubulin and MTs (high-frequency): $\kappa = 8.41 $. High-frequency polarizability of tubulin: $\alpha = 2.1 \times 10^{-33}\, \rm C\, m^{2}/V.$ Sources include both theoretical predictions and experimental results across cryogenic and biological regimes. We also note the finding of electrical oscillations of bundles of microtubules in the brain  \cite{BrainBundleMT}.}

\begin{tabular}{|p{5cm}|c|p{2cm}|p{6cm}|}
\hline
\textbf{Parameter} & \textbf{Value} & \textbf{Units} & \textbf{Citation / Note} \\
\hline
Thermal energy $k_B T$ at $T = 310^{\rm o}\,\mathrm{K}$ & $4.3 \times 10^{-21}$ &   J &\\
                              & $\sim 27$  \color{black}& meV & \\
                              & $\sim 15.7$ \color{red}  \color{black}& $k_B T$ units &  \\
\hline
GTP→GDP hydrolysis energy & $\sim 0.42$ & eV & \cite{towardsTests} \\
\hline
Dipole moment of tubulin dimer (GTP state) & $\sim 1700$ & Debye & $\approx 5.7 \times 10^{-27}$\,C$\cdot$m\cite{tuszynski2014} \\
Dipole moment angle change (GTP→GDP) & $\sim 27^\circ$ & degrees & Conformational shift upon hydrolysis\cite{towardsTests} \\
\hline
Microtubule protofilaments & $13$ & unitless & Standard MT structure~\cite{towardsTests} \\
Helical pitch & $\sim 12$ & nm per turn & \\
Outer diameter & $25$ & nm & \cite{towardsTests} \\
Inner lumen diameter & $14$–$15$ & nm & \\
\hline
Tubulin monomer mass & $\sim 50$ & kDa & $\alpha$ or $\beta$ subunit\cite{tuszynski2014} \\
Tubulin dimer mass & $\sim 100$ & kDa & Heterodimer \\
\hline
Typical microtubule length (neuronal) & $1$–$100$ & µm & Varies with cell type\cite{towardsTests} \\
\hline
Cryptochrome decoherence time & $> 10^{-5}$ & s & Predicted, bird magnetoreception~\cite{gauger2011} \\
Photosystem I coherence time & $300$–$800$ & fs & \cite{engel2007} \\
Light harvesting complex (FMO) coherence & $> 300$ & fs & Room temp evidence~\cite{engel2007} \\
Photosynthetic quantum efficiency & $\sim 0.95$ & unitless & \cite{engel2007} \\
\hline
Soliton propagation speed (microtubules) & $2$–$20$ & m/s & Nonlinear excitations~\cite{towardsTests} \\
Time for soliton to travel 1 µm & $50$–$500$ & ns & $\tau = \frac{1\,\mu m}{v}$ \\
\hline
SPR evanescent field strength & $10^4$–$10^5$ & V/m & Estimated for optical range~\cite{tuszynski2014} \\
Dipole-field coupling energy & $\sim 10^{-21}$ & J & $\Delta E = -\vec{p} \cdot \vec{E}$, assuming $p \sim 10^{-27}$\,C$\cdot$m \\
\hline
Surface plasmon group velocity & $\sim 10^7$ & m/s & Guided modes in nanofilms \\
Travel time across 1 µm & $\sim 0.1$ & ps & $\tau = \frac{1\,\mu m}{v}$ \\
\hline
Physiological pH (neurons) & $7.2$–$7.4$ & pH & \cite{towardsTests} \\ \hline 
Physiological salinity & $\sim 150$ & mM NaCl & \\ \hline
Effective dielectric constant $\varepsilon/\varepsilon_0$ (MT interior) & $10$ $-$ $40$  & unitless &   Depends on hydration  and polarization~\cite{tuszynski2014}   
\\
\hline
\end{tabular}
\label{tab:bioquantum_parameters}
\end{table*}

To quantify the feasibility of such operations, we estimate the strength of interaction between the tubulin dipole and external fields. Referring to tables \ref{tab:parameters} and \ref{tab:bioquantum_parameters}: for surface plasmons launched on nanostructured gold substrates, the evanescent field at a distance of $100$--$200\,\mathrm{nm}$ into the aqueous medium can reach magnitudes of $E_{\mathrm{plasmon}} \sim 10^5\,\mathrm{V/m}$. Using a typical tubulin dipole moment $p \approx 1700\,\mathrm{D} \approx 5.67 \times 10^{-27}\,\mathrm{C{\cdot}m}$ \cite{tuszynski2014}, the resulting interaction energy is
\begin{equation}
    \Delta E_{\mathrm{plasmon}} = p E_{\mathrm{plasmon}} \approx 0.35\,\mathrm{meV}.
\end{equation}

By contrast, during a neuronal action potential, the membrane depolarizes by approximately $\Delta V_{\mathrm{mem}} \sim 0.1\,\mathrm{V}$. At a distance of $r \sim 500\,\mathrm{nm}$ into the axon, where microtubule bundles typically reside, the induced radial electric field is
\begin{equation}
    E_{\mathrm{axon}} = \frac{\Delta V_{\mathrm{mem}}}{r} \approx 2 \times 10^5\,\mathrm{V/m},
\end{equation}
leading to a coupling energy of
\begin{equation}
    \Delta E_{\mathrm{axon}} = p E_{\mathrm{axon}} \approx 0.71\,\mathrm{meV}.
\end{equation}
Both values fall within the sub-meV regime and are below thermal fluctuations at $T = 310\,\mathrm{K}$ ($k_{\mathrm{B}}T \approx 27\,\mathrm{meV}$),  and far below the energy associated with GTP to GDP hydrolysis ($\simeq 317~\rm meV$).Notably, the coupling energy from action potentials is approximately twice that of the lowest plasmonic interaction. 

These values are $2$--$3$ orders of magnitude smaller than $k_B T$, implying that such weak fields alone are insufficient to induce dipole state transitions in thermal equilibrium unless enhanced local fields are used such as those created by defects or nanostructures at a conducting surface. Even without such assistance (which can increase the field strength by several orders of magnitude locally), in the structured environment of the microtubule interior—particularly when interacting with ordered water and neighbouring dipoles—these couplings may contribute to the initiation or modulation of coherent quantum dynamics. In this sense, field-driven transitions may act as subtle biasing agents rather than direct triggers of quantum state evolution.   

This suggests that biological activity may induce or modulate quantum transitions in MT systems via electric field-driven mechanisms~\cite{lioub}, but only upon enhancement to bring the interaction above the competing thermal bath, while the plasmonic field provides a means to optically interrogate or stimulate such quantum states without having to use living cells (action potentials' effects can be conceivably replicated using surface-mounted MT networks addressed by fields amplified by nanostructures) ~\cite{kim}. An analogy of  trees, branches and leaves springs to mind here: while sustained wind (hydrolysis and large energy dissipation events)  may create waves in branches or even whole trunks, gusts can also be registered  as "ripples" of much faster dynamics and lower energies can still be seen as ``ripples" that appear and move fast on  -by comparison- slower-moving and wider-amplitude waves. 

The close equivalence of field strengths from these two disparate sources underscores the dual utility of MT bundles when considering substrates for the basic science as well as the specific application to scalable, ambient temperature "wet" quantum computation: they are both responsive to endogenous bioelectrical dynamics and accessible to engineered photonic quantum probes albeit with field-enhancers necessary such as the proposed surface plasmon entanglement transduction system \cite{QuantumLifeCh7,Altewischer2002}, where dipole transitions in tubulin could be modulated and read via optical coherence measurements.

We conclude that MT networks, with dipolar and solitonic degrees of freedom embedded in a well-defined lattice, can plausibly perform elementary quantum operations but the requirement remains for  decoherence  to be sufficiently delayed -as is here seen possible by structural or environmental isolation.

\begin{figure}[ht]
\begin{center}
\includegraphics[width=12cm]{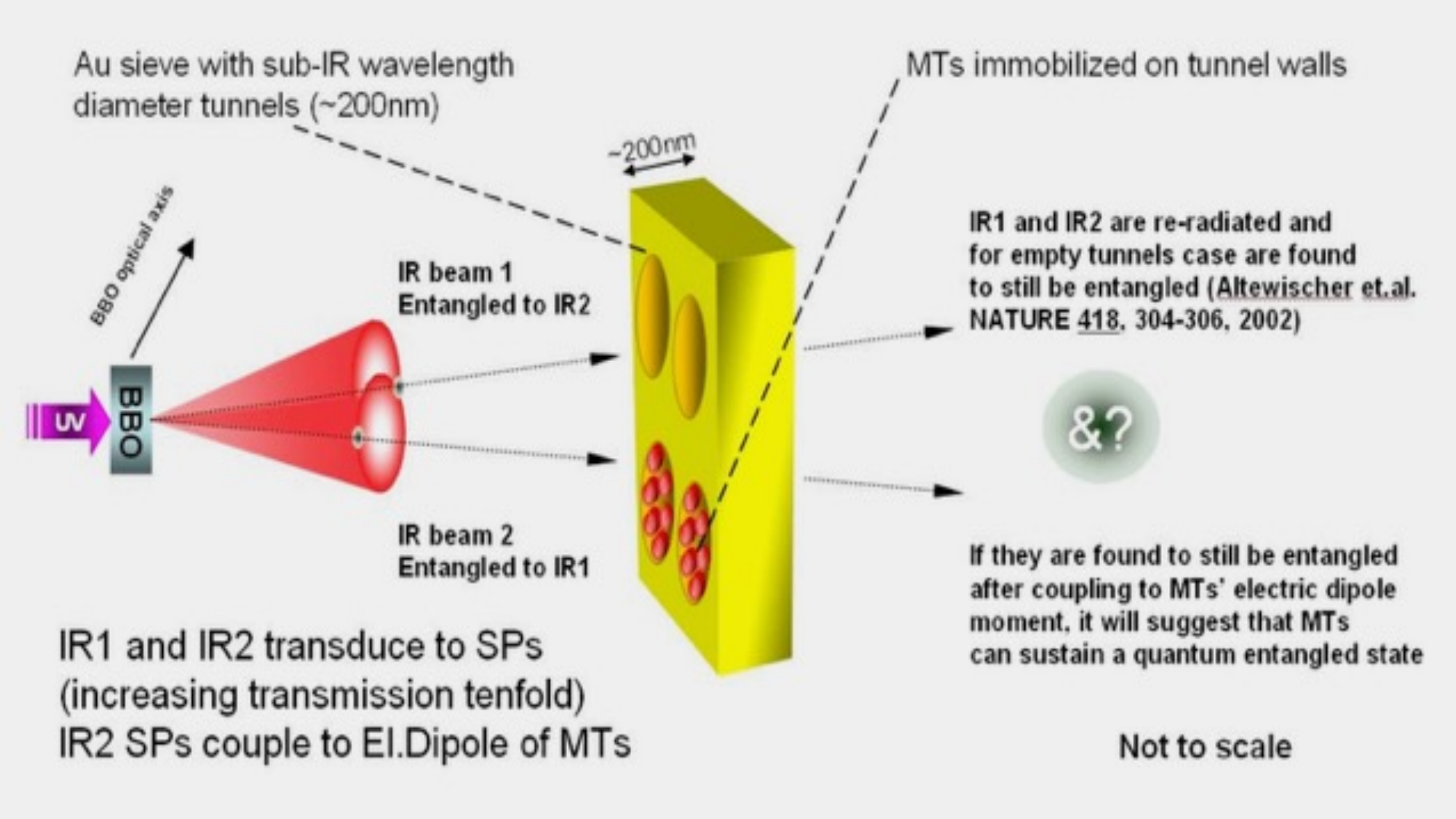}
\end{center}
\caption{A photonic entanglement transduction system, in which entangled photons are converted 
sequentially into surface plasmons, and then couple to protein dipole states. Representation is not to scale, and when implemented as a perforated conducting screen, it would be hundreds of tunnels per illuminated spot instead of just the four shown here. Figure taken from \cite{QuantumLifeCh7}. }
\label{fig:plasmons}
\end{figure}

Previously published simulations suggest that, under physiologically plausible parameters (pH, dipole strength), tubulin may support coherence times on the order of microseconds~\cite{Craddock2014}. To probe this experimentally, we have proposed~\cite{{QuantumLifeCh7}} a photonic entanglement transduction system, where entangled photons are sequentially converted to surface plasmons, as has been so eloquently accomplished by Altewischer {\it et al.} in 2002~\cite{Altewischer2002}, and then couple to protein dipole states, following principles established in mesoscopic plasmonics~\cite{Oberparleiter2000} (see Figure \ref{fig:plasmons}).

Our case is agnostic to the details of the entangled photon pairs generation scheme, but the well-established type-II phase-matched spontaneous parametric down-conversion in a $\beta$-barium borate crystal can serve as exemplar. Such an arrangement would be producing Einstein-Podolsky-Rosen-correlated infrared photons with polarization entanglement described by the state
\begin{align}
|\Psi\rangle = \frac{1}{\sqrt{2}}\left(|\leftrightarrow\rangle_1 |\updownarrow\rangle_2 + e^{i\alpha}|\updownarrow\rangle_1 |\leftrightarrow\rangle_2\right)
\end{align}
where the phase $\alpha$ is tunable via crystal orientation or additional birefringent elements~\cite{Kwiat1995}.

One photon of each entangled pair is directed into a plasmonic interface based on the architecture of Altewischer \textit{et al.}~\cite{Altewischer2002}, wherein photon-to-plasmon conversion occurs via subwavelength apertures in a gold film Figure 2 of  \cite{QuantumLifeCh7} reproduced here. To couple plasmons to biomolecules, it is possible to modify this setup by coating the perforations with a monolayer of immobilized tubulin dimers~\cite{towardsTests} that can be assembled into full MTs. The evanescent fields of the plasmons interact with the permanent dipole moments of the proteins, potentially transferring entanglement into molecular dipole degrees of freedom but this transfer has never been documented by experiment before.  The pertinent Coulombic interactions are described in section \ref{sec:qucom}, see \eqref{dimerenv}. For typical values of the permittivity $\varepsilon \sim 80$, the analysis of \cite{Xiang} shows that such interactions yield the short decoherence time of \eqref{fs} for a tubulin dimer. Here we see this as low-hanging fruit because following interaction, the plasmons that are reconverted to photons can be tested for residual entanglement with their twin photons using quantum state tomography and observation of partial or full entanglement would be strongly suggestive of coherent information transfer between light and the protein dipole system. There should be two qualitatively and quantitatively different peaks associated with the re-emission of entangled photons: those that only interacted with the metal surface and those that are emitted from the MTs via the metal surface. Prior experiments by our group confirmed the feasibility of tubulin immobilization and its optical response using surface plasmon resonance (SPR) and refractometry. We reported \cite{Biosystems2004} concentration-dependent shifts in refractive index and dielectric constant of
\begin{align}
\frac{\Delta n}{\Delta c} &= (2.0 \pm 0.5) \times 10^{-3} \,\text{ml/mg}\,,\nonumber \\
\frac{\Delta \varepsilon}{\Delta c} &= (0.5 \pm 0.1) \times 10^{-3} \,\text{ml/mg}\,,
\end{align}
in agreement with direct refractometric measurements ~\cite{JModOptics2003} .

Our own simulation ~\cite{Biosystems2004} yielded dipole moments of 552~D and 1193~D  for $\alpha$- and $\beta$-monomers respectively, and 1740~D for the dimer, with polarizability $2.1 \times 10^{-33} \, \mathrm{C\,m^2/V}$, high-frequency dielectric constant $\varepsilon_r = 8.41$ and refractive index $n = 2.90$ (which was also experimentally confirmed at 527~nm by label-free imaging of cytoskeleton of living mammalian cells by an independent group \cite{Biophys2014J}). These values suggest strong coupling potential to plasmonic near-fields, making tubulin a viable candidate for quantum optical probing~\cite{Tuszynski1998}, and additional experimentally determined values of interest are listed in Table \ref{tab:bioquantum_parameters}. 

By varying the plasmonic path length (200--800~nm), it should be possible to measure decoherence times and assess the persistence of entanglement post-interaction, thereby testing the viability of protein-based ``bioqubits" in biological quantum computing contexts \color{black} (see Table \ref{tab:benchmarks_sectionV}).\color{black}

Finally before closing this subsection, we
mention one more experimental path, which is not dissimilar to the aforementioned RNA memory transplant in snails~\cite{rnamemory}, mentioned in footnote~\ref{foot1}.
Indeed, it is known that Drosophila can form stable olfactory memories of isotopically distinct molecules—such as deuterated odorants—in ways suggestive of vibrational encoding~\cite{MershinFlies2011,Ball}, and that overexpression of tau microtubule-associated protein in their memory-encoding and microtuble-rich mushroom bodies disrupts normal memory formation and retrieval~\cite{MershinFlies2004}. Such a behaviour is consistent with the QED-cavity models of MT function~\cite{mn1,mmn1} and is not predicted by conventional biological frameworks, in which MTs are seen as primarily structural and cell-mobility elements of the cytoskeleton, not typically seen as closely involved with learning and memory. We cannot help mentioning here that the robust findings of \cite{rnamemory}, according to which the RNA extracted from trained Aplysia can transfer memory-induced behavior to naïve individuals, points towards a further exploration of how such RNA molecules might engage with the MT network—along which they are known to travel, affecting synaptic formation~\cite{Holt}—as that would be one possible coupling route between biochemical and biophysical substrates of cellular memory.
The alert reader should recall here that a known interaction of RNA molecules with MTs occurs in cell mitosis (see discussion in the first reference of \cite{nano}).

\section{Conclusions and Outlook}\label{sec:concl}

By integrating QED-based models of microtubule (MT) dynamics~\cite{mn1, mmn1, nem} with principles of quantum information theory, and by employing the best available physicochemical parameters, we conclude that MTs can, in principle, function as multi-level qudit processors, provided that coherent quantum states of tubulin dimers can be controllably prepared (see  \cite{towardsTests}). 

Building upon earlier representations~\cite{mexico} of classical MT tubulin dipole moments as a pseudospin non-linear $\sigma$-model, which admits solitonic solutions ranging from kinks and spikes to snoidal and helicoidal waves, we have argued that these solitons can be interpreted as coherent quantum pointer states of tubulin dimer-dipole excitations. In this construction, the fundamental hexagonal unit of the honeycomb lattice representing tubulin dipole arrangements (see Fig.~\ref{MT}) corresponds to the unit of quantum information storage, explicitly: a quDit. Unlike conventional approaches, in our model the quDit is not defined by the $\alpha$ and $\beta$ dipole conformations alone, but by combinations of four dipole quantum states associated with the parallelogram formed by four of the seven dimers in the fundamental hexagonal cell (see Fig.~\ref{MT1b}). 

External stimuli drive quantum fluctuations and entanglement in this fundamental unit, leading—within the decoherence time of the MT—to a ``decision'' on optimal pathways for signal and energy transport via the formation of solitonic states. Double-helix snoidal waves, for instance, arise from the incomplete collapse of tubulin dimer quanta into pointer states~\cite{zurek2} and propagate information along individual MTs and across MT networks in a dissipation-free manner.\footnote{It is worth noting that snoidal waves, here proposed as information carriers in MTs, also emerge in the dynamics of the pendulum. This parallel has long underpinned timekeeping devices, such as turret clocks, which translate pendular oscillations into the motion of clock hands. In analogy, snoidal solitons in MTs convey the passage of time and information within cellular networks. The shared mathematics of pendulum motion and MT dynamics underscores this analogy, while the connections between MTs mediated by MAPs ({\it cf.}~Fig.~\ref{gates}) echo the mechanical linkages that allow turret clocks to operate as logic devices.}

We have also discussed a QED cavity model for MTs~\cite{mn1,nem}, wherein the ordered water in the lumen acts as a high-quality electromagnetic cavity, bounded by tubulin dipole quanta. Strong interactions between the dimer dipoles and the ordered-water dipoles near the walls yield a highly isolated cavity, with environmental coupling primarily due to leakage of water dipole quanta through the imperfect protein walls. As shown in~\cite{mn1}, this configuration supports decoherence times on the order of $10^{-6}$~s, allowing micron-scale MTs to remain quantum-wired (entangled) and capable of dissipation-free energy and signal transport.

Absent this cavity effect, decoherence times for individual tubulin dimers—limited by direct dipole–dipole interactions—would be much shorter, $\mathcal{O}(1$–$100)$~fs (see Eq.~\eqref{fs}). While such times suffice for quantum wiring across $\sim 40$~\AA\ (adequate for entire photosynthetic antenna complexes~\cite{algae,collini}), they are insufficient for coherent wiring of full MTs or MT networks. Thus, the cavity mechanism is essential to extend coherence times to biologically relevant scales.

Taken together, these considerations outline the requirements for biocomputation via MT networks at ambient temperatures, as introduced in Sec.~\ref{sec:qucom}~\cite{critique}. Specifically, we have provided:  
(i) a precise description of the quDit states (dimension $D=4$ in the simplest case);  
(ii) a mechanism for entanglement of these states, including the measurement basis; and  
(iii) a pathway for maintaining coherence on the required timescale to support quantum wiring.  

Within this framework, quDit-gate operations arise naturally from dipole–dipole and dipole–field interactions, while solitonic excitations enable coherent, low-loss state transfer. MT networks, therefore, present a biologically plausible platform for quantum information processing (Fig.~\ref{gates}), and their experimental interrogation could advance prospects for scalable quantum computation.

For reference, Tables~\ref{tab:parameters} and~\ref{tab:bioquantum_parameters} summarize typical parameter values relevant to proposed quantum information mechanisms in biosystems, compared with those in our MT cavity model. Ultimately, determining whether MTs can serve as substrates for biocomputation requires targeted experiments of the type proposed here and in related studies. A complementary approach is to design synthetic quantum devices inspired by MT architectures, for example by using engineered spin systems to replicate their essential features.

\color{black}
Before closing, we would like to mention several experimental directions. There are research works dealing with electron transport modeling in MT~\cite{etcomp}. There are also studies on the role of MT as regulators of the shape and functions of network of neurons in the brain~\cite{MTnn}, as well as research works on the electrical behaviour of (bovine) MT (viewed as complex electrical networks) and its potential connection with the role of MT as a medium for evolutionary computation~\cite{elbehMT}. Moreover, there have been analyses on classical computing modellng of MT networks, attempting to understand potential links with their physiology, in particular reproduction of features related to degenerating neurons in disease states, such as Alzheimer's disease ~\cite{MTphysio}. There have also been efforts aimed at the construction of  an electrical analogue computer out of  MT protofilaments~\cite{analogue}, where using operational amplifiers, capacitors, and resistors, the authors designed analytically the bioelectronic circuit of the MT protofilament, in an attempt to understand computational aspects of these biological entities. Information transfer and storage in brain has also been studied from an MT perspective  \cite{MTinfoproc}, considering them as a communication channel. Although the above works deal with classical computational aspects of MTs, nonetheless we may think of combining such attempts with our studies here, which make use of the quantum computing aspects of MT, viewing them as QED cavities.
\color{black}

{\it Affaire \`a suivre...}


\section*{Acknowledgments}

The work of N.E.M. is supported in part by the UK Science and Technology
Facilities research Council (STFC) under
the research grant  ST/X000753/1. 
NEM also acknowledges participation in the COST Association Actions CA21136 ``Addressing observational tensions in cosmology with systematics and fundamental physics (CosmoVerse)” and CA23130 ``Bridging high and low energies in search of quantum gravity (BridgeQG)".
AM wishes to acknowledge the support received from www.RealNose.ai and the intellectual boost received from participants of the www.OsmoCosm.org MIT conferences and the MIT IAP class ``Making Sense of Scent".
DVN would like to thank his family Olga and Odysseas for encouragement and patience 
and the Digital Health  Literacy \& Policy Hub Foundation (digitalhealth-hub.com) for support.

\end{document}